\documentclass{aa}

\newcommand{\bs}{$\langle B_{\rm s}\rangle$}

\usepackage[colorlinks=true,     linkcolor=blue, citecolor=blue, filecolor=blue, urlcolor=blue]{hyperref}
\usepackage{graphicx}
\usepackage{txfonts}
\begin{document}

   \title{Spectrum and energy levels of the high-lying singly excited configurations of Nd~III\thanks{All tables are available in full at the CDS.}}

   \subtitle{New Nd III experimental energy levels and wavelengths, and transition probability and ionisation energy calculations}

   \author{M. Ding\inst{1}
          \and
          A. N. Ryabtsev\inst{2}
          \and
          E. Y. Kononov\inst{2}
          \and
          T. Ryabchikova\inst{3}
          \and
          J. C. Pickering\inst{1}%\fnmsep\thanks{Just to show the usageof the elements in the author field}
          }

   \institute{Physics Department, Imperial College London, Prince Consort Road, London, SW7 2AZ, UK \\
              \email{milan.ding15@imperial.ac.uk}
            \and
             Institute of Spectroscopy, Russian Academy of Sciences, Troitsk, Moscow, 108840, Russia
             \and
             Institute of Astronomy, Russian Academy of Sciences, Pyatnitskaya 48, Moscow 119017, Russia
             %\thanks{The university of heaven temporarily does notaccept e-mails}
             }

   \date{Received ?; accepted ?}

% \abstract{}{}{}{}{} 
% 5 {} token are mandatory
 
  \abstract
  % context heading (optional)
  % {} leave it empty if necessary  
   {}
  % aims heading (mandatory)
   {To accurately determine bound-to-bound transition wavelengths and energy levels of the high-lying open-shell configurations 4f$^3$7s, 4f$^3$6d, and 4f$^3$5f of doubly-ionised neodymium (Nd~III, $Z=60$) through high-resolution spectroscopy and semi-empirical calculations.}
  % methods heading (mandatory)
   {Fourier transform spectra of Nd Penning and hollow cathode discharge lamps were recorded within the region 32\,500--54\,000~cm$^{-1}$ (3077--1852~\AA) and grating spectra of Nd vacuum sliding sparks were recorded within the regions 820--1159~\AA{} and 1600--3250~\AA{}. New energy levels were found using the observed wavelengths measured accurate to a few parts in $10^8$ in Fourier transform spectra and to a few parts in $10^7$ in grating spectra. Atomic structure and transition probability calculations of Nd~III were made using the Cowan codes by adjusting energy parameters to fit all known Nd~III levels. Nd-rich stellar spectra were also used to evaluate the new calculations.}
  % results heading (mandatory)
   {In total, 355 transitions were classified from observed spectra involving 116 previously experimentally unknown energy levels of the 4f$^3$7s, 4f$^3$6d, and 4f$^3$5f configurations of Nd~III, all reported here for the first time. One newly identified level of the 4f$^3$5d configuration is also reported. Typical level energy uncertainties are 0.01~cm$^{-1}$ for the 4f$^3$7s and 4f$^3$6d levels and 0.3~cm$^{-1}$ for the 4f$^3$5f levels. In addition, calculated energy levels up to 130\,936~cm$^{-1}$ are presented, including eigenvector composition and calculated level lifetimes. Calculated transition probabilities and wavelengths between 1900--50\,000~\AA{} are also presented. Using newly established levels of the 4f$^3$7s configuration and the recently established levels of the 4f$^3$6s configuration, the ionisation energy of Nd~III was estimated at $178\,090\pm330$~cm$^{-1}$, doubling the accuracy of the previously published value.}
  % conclusions heading (optional), leave it empty if necessary 
   {These results will enable more accurate analyses of astrophysical plasmas involving the Nd~III ion, such as in chemically peculiar stars and kilonovae. Results from the Cowan code calculations will also aid the identification of Nd~III transitions unobserved in the laboratory spectra of this work and support transition probability measurements of Nd~III.}

   \keywords{Atomic data -- Line: identification -- Methods: laboratory: atomic -- Methods: data analysis -- Stars: abundances -- Stars: chemically peculiar
               }

   \maketitle
%
%-------------------------------------------------------------------
\nolinenumbers
\section{Introduction} \label{sec:intro}
Accurate and complete knowledge of atomic spectra of the lanthanide elements ($Z=57-71$) is necessary for a variety of astronomical applications. For example, in reliable astrophysical chemical abundance analyses in spectroscopic studies of chemically peculiar stars \citep[e.g.,][]{przybylski1977iron, cowley2000abundances, ryabchikova2006rare} and their pulsational wave propagations \citep[e.g.,][]{savanov1999radial}. Wavelengths of the spectral lines used in these analyses must be known with accuracies matching or exceeding those provided by the existing and planned high-resolution modern ground and space-based telescopes with resolving powers of up to 10$^5$. Importantly, in disentangling the spectra of kilonovae \citep[e.g.,][]{kasen2013opacities, tanaka2013radiative, smartt2017kilonova, tanaka2018properties, watson2019identification, gaigalas2019extended, tanaka2020systematic, even2020composition, cowan2021origin, domoto2022lanthanide}, the spectral data of the lanthanide elements of the first few ionisation stages also need to be as complete and as accurate as possible. 

Experimental investigation of the atomic structure of the lanthanide elements using high-resolution spectroscopy is a key step in addressing the need for complete and accurate knowledge of their spectral data. Not only do the measured transition wavelengths fulfill the accuracy requirements, but the experimentally established energy levels also allow accurate transition probability (TP) measurements or calculations to be undertaken, e.g., in the present work, semi-empirical calculations were made from adjusting energy parameters of the Cowan codes \citep{cowan1981theory, kramida2021suite} to optimise prediction of the level energies and eigenvector compositions that are used for calculations of theoretical TPs.

The original experimental investigations of Nd~III atomic structure were by \cite{dieke1961emission} and \cite{dieke1963spectra}, which were later extended by \cite{aldenius2001}, \cite{ryabchikova2006rare}, and \cite{ding2023spectrum}. This paper reports the continuation and extension of the work of \cite{ding2023spectrum} that had reported the classification of 623 transitions of 105 new and 39 improved energy levels of the 4f$^4$, 4f$^3$5d, 4f$^3$6s, and 4f$^3$6p configurations of Nd~III. In the present work, 355 additional transitions of Nd~III are further classified by Fourier transform (FT) and grating spectroscopy of Nd discharge sources, which determine 116 new energy levels of the 4f$^3$7s, 4f$^3$6d, and 4f$^3$5f configurations for the first time. The further analysis of stellar absorption spectra of Nd~III has also led to the identification of one new 4f$^3$5d level, increasing the total number of known levels of Nd~III now to 261.

All level energies of the 4f$^3$7s and 4f$^3$6d configurations are determined by transition wavenumbers measured by FT spectroscopy of a Nd Penning discharge lamp (PDL) \citep[e.g.,][]{finley1979continuous, heise1994radiometric} within the region 32\,500--54\,000~cm$^{-1}$ (3077--1852~\AA), accurate to a few parts in 10$^8$. All level energies of the 4f$^3$5f configuration are determined by grating spectroscopy of Nd vacuum sliding spark (VS) discharges within the region 820--1159~\AA{}, accurate to a few parts in 10$^7$. %The strongest 4f$^3$6d -- 4f$^3$5f transitions were unobserved due to spectral range limitations of the instruments, but their Ritz wavenumbers are established. %The present analysis of the 4f$^3$7s and 4f$^3$6d configurations was enabled equally by FT and grating spectroscopy of Nd; the small number of 4f$^3$6p -- 4f$^3$7s and 4f$^3$6p -- 4f$^3$6d transitions observed in the FT spectra used for wavenumber measurements were only classifiable due to the existence of many weaker lines of these transition arrays in the grating spectra.

The ionisation energy is one of the fundamental constants of an ion, which is also important for modelling astrophysical plasma conditions. Adopting the method from \cite{sugar1973ionization}, the Nd~III ionisation energy is estimated to be $178\,070\pm330$~cm$^{-1}$ using previously unknown experimental level energies.

%--------------------------------------------------------------------
\section{Experimental details}\label{sec:experiment}
Full experimental details of the majority of the atomic emission spectra analysed in the present work are in \cite{ding2024thesis} and \cite{ding2023spectrum}, therefore only a summary is provided here. Prior to wavelength and relative intensity calibration, the wavelengths and relative intensities of each spectral line were obtained by fitting the spectra using model line profiles, and resulting line parameters were tabulated as line lists for the subsequent energy level analysis. %The relative intensities for all observed spectral lines in this work are recommended as a guide only, as the experiments were not designed for accurate branching fraction measurements.

\subsection{Fourier transform spectroscopy of Nd}
The FT spectra used in this work were the two Nd-Ar PDL spectra measured in and labelled as E and F in Table 2 of \cite{ding2023spectrum}, covering the regions 32\,480--44\,422~cm$^{-1}$ (3079--2251~\AA) and 44\,422--53\,822~cm$^{-1}$ (2251--1858\AA), respectively. These were recorded on the high-resolution f/25 Imperial College VUV FT spectrometer \citep{thorne1987fourier}, at resolutions 0.07 and 0.08~cm$^{-1}$ respectively, which were chosen such that the resolving power was limited by the Doppler width of the PDL spectral lines. The PDL cathodes were 99.5\% pure Nd and run with argon carrier gas at a current of 750~mA for both spectra and at pressures of 1.7$\times10^{-3}$ and 2.0$\times10^{-3}$~mbar respectively. The observed wavenumbers were calibrated using Ar~II standard lines recommended by \cite{learner1988wavelength} and measured by \cite{whaling1995argon}. The wavenumber uncertainties of non-blended, non-self-absorbed lines with signal-to-noise ratio (S/N) greater than 100 were no more than 0.003~cm${-1}$ ($\sim$ 0.0003~\AA{} at 3000~\AA{}). The spectra were intensity calibrated to a relative scale using a deuterium standard lamp. Two Nd-Ar hollow cathode lamp (HCL) spectra had also been recorded in exactly the same spectral regions in \cite{ding2023spectrum}. The 4f$^3$6p -- 4f$^3$7s and 4f$^3$6p -- 4f$^3$6d transitions used in the present work to find the 4f$^3$7s and 4f$^3$6d levels were notably not observed in these two HCL spectra, and this aided their classifications as one would expect transitions from higher-lying levels to be lower in intensity in the hollow cathode lamp due to its lower effective temperature and hence correspondingly lower level populations for higher-lying levels compared to the PDL.

\subsection{Grating spectroscopy of Nd}
The grating spectra used in this work include all four Nd VS spectra presented in Table 4 of \cite{ding2023spectrum}, and these were recorded at VS currents up to 1500~A. Two of these spectra were recorded using the 6.65~m normal incidence spectrograph at the Institute of Spectroscopy in Troitsk using photographic plates, spanning the regions 390--1525~\AA{} and 1600--2536~\AA{}. The other two spectra were recorded using the 10~m normal incidence spectrograph at the National Institude of Standards and Technology (NIST), covering 2330--3250~\AA{}, also using photographic plates. Impurity lines of oxygen, carbon, nitrogen, and silicon in various ionisation stages \citep{kramida2022nist}, as well as lines of Nd~IV \citep{wyart2007analysis} and Nd~V \citep{meftah2008spectrum}, were used for wavelength calibration in the range 390--1525~\AA{}. The other grating spectra were wavelength calibrated using Nd III Ritz wavelengths determined in \cite{ding2023spectrum}. Relative intensities were calibrated using approximate plate response curves and sensitivities. The wavelength uncertainty was estimated at $\pm0.006$~\AA{} using the RMS deviations of wavelengths from corresponding reference wavelengths. Due to the much higher currents of the VS discharges, the weaker transitions from high-lying Nd~III levels that were either around the noise level or absent in the Nd-Ar PDL FT spectra were of moderate intensity in the Nd VS grating spectra.

Additionally, a line list of a grating spectrum in the range 820--1159~\AA{} was used \citep{wyart2006private}. This spectrum was recorded using a VS in the sliding mode of operation at 200~A current on the NIST 10-m normal incidence spectrograph. The wavelength uncertainty of this spectrum was estimated at $\pm0.006$~\AA{} using the RMS of differences between wavelengths of Nd~III lines in this spectrum and the spectrum from Troitsk in the range 390--1525~\AA{}. The Troitsk Nd VS spectrum was taken at 1500~A and has a much higher background, many weaker lines were not observed and some lines were also blended with the second-order lines with shorter wavelengths. Therefore, the additional line list of \cite{wyart2006private} in the range 820--1159~\AA{} was used for wavelength measurements of the 4f$^3$5d – 4f$^3$5f transitions.

\section{Empirical spectrum analysis of Nd III}
\subsection{Methodology}
The empirical spectrum analysis of an atom or ion involves assigning upper and lower energy levels, i.e., a pair of energies and a pair of $J$ values, to each of its transitions observed in the laboratory. This can be done by matching observed wavelengths of spectral lines with theoretical level energy separations, and by matching observed relative intensities of spectral lines with theoretical TPs and expected relative level populations. Such an analysis requires information on the spectrum of atomic transitions to be extracted into line lists by fitting experimental spectra using model line profiles. A line list contains the important parameters for each line, namely the wavenumber and relative intensity and their uncertainties. Other parameters such as S/N and FWHM are also a part of the line lists as these can hint at potential problems with the fitted lines, such as unreliable weak and blended lines.

The matching of intensity patterns and level energy values to find new energy levels can be ambiguous in complex atomic spectra where the number of spurious candidate energy levels increases rapidly with increasing number of lines and increasing experimental and theoretical uncertainties in energies and relative intensities. Extra observational evidence is often required for correct line classifications. The methodology and strategies employed for the analysis of Nd~III are discussed in detail in \cite{ding2023spectrum}.
\subsection{Results}
\subsubsection{Energy levels}\label{sec:energy_levels}
All newly found level energies of the 4f$^3$7s and 4f$^3$6d configurations were optimised by inputting wavenumbers of 579 Nd~III lines fitted from the Nd-Ar PDL FT spectra \citep{ding2023spectrum} into the computer program LOPT by \cite{kramida2011program}, where level energy uncertainties, Ritz wavenumbers, and Ritz wavenumber uncertainties were also estimated. All newly found 4f$^3$5f level energies were estimated using transition wavelengths recorded by \cite{wyart2006private} in the range 820--1159~\AA{} and input into LOPT with the connecting 4f$^3$5d levels at fixed energies. The differences between fixing and not fixing the 4f$^3$5d level energies are negligible due to the much higher uncertainties of the 4f$^3$5d -- 4f$^3$5f grating wavelengths. This fixing of 4f$^3$5d levels was chosen as the higher-resolution FT measurements used to find the 4f$^3$5d levels in \cite{ding2023spectrum} are more reliable, i.e., higher accuracy and better-resolved blends. 

The 116 newly identified energy levels of the 4f$^3$7s, 4f$^3$6d, and 4f$^3$5f configurations of Nd~III are included in Table \ref{tab:nd_levels}.
{\setlength{\tabcolsep}{4pt}
\begin{table*}[!th]
\footnotesize
\renewcommand{\arraystretch}{1.3}
\caption{Energy levels experimentally established and semi-empirically calculated for Nd~III (extract). \label{tab:nd_levels}}
\centering
\begin{tabular}{ccccccclll}
\hline\hline
\multicolumn{1}{c}{Par.} &
\multicolumn{1}{c}{Label} &
\multicolumn{1}{c}{$J$} &
\multicolumn{1}{c}{$E_e$} &
\multicolumn{1}{c}{$\Delta E$}&
\multicolumn{1}{c}{$\tau$}&
\multicolumn{1}{c}{$g^l$}&
\multicolumn{3}{c}{Eigenvector Composition}
\\
\multicolumn{1}{c}{} & 
\multicolumn{1}{c}{} & 
\multicolumn{1}{c}{} & 
\multicolumn{1}{c}{(cm$^{-1}$)}&
\multicolumn{1}{c}{(cm$^{-1}$)}&
\multicolumn{1}{c}{(ns)}&
\multicolumn{1}{c}{} & 
\multicolumn{3}{c}{(\%)}\\
\multicolumn{1}{c}{(1)} & 
\multicolumn{1}{c}{(2)} & 
\multicolumn{1}{c}{(3)} & 
\multicolumn{1}{c}{(4)}&
\multicolumn{1}{c}{(5)}&
\multicolumn{1}{c}{(6)}&
\multicolumn{1}{c}{(7)} & 
\multicolumn{3}{c}{(8)}\\
\hline
e&4f$^4$ $^5$I                  & 4 & 0.000(*)       &   4.2 & & 0.604 & 97 4f$^4$ $^5$I & & \\
\multicolumn{1}{c}{\vdots} & \multicolumn{1}{c}{\vdots} & \multicolumn{1}{c}{\vdots} & \multicolumn{1}{c}{\vdots} & \multicolumn{1}{c}{\vdots} & \multicolumn{1}{c}{\vdots} & \multicolumn{1}{c}{\vdots} & \multicolumn{1}{c}{\vdots} & \multicolumn{1}{c}{\vdots} \\
e&4f$^3$($^4$I$^\circ$)5f $^5$L         & 6 & 114\,016.4(6)    &  -4.8 & $5.2\times10^{-10}$ & 0.760 & 79 4f$^3$($^4$I$^\circ$)5f $^5$L  &    13 4f$^3$($^4$I$^\circ$)5f $^5$K  & \;\:2 4f$^3$($^2$H2$^\circ$)5f 3K \\
e&4f$^3$($^4$I$^\circ$)5f $^5$M	        & 7 & 114\,015.1(5)    &  -6.5 &  $5.1\times10^{-10}$ & 0.768 & 88 4f$^3$($^4$I$^\circ$)5f $^5$M  & \;\:7 4f$^3$($^4$I$^\circ$)5f $^5$L  & \;\:3 4f$^3$($^2$H2$^\circ$)5f 3L \\
e&4f$^3$($^4$I$^\circ$)5f $^5$K         & 5 & 114\,021.3(4)    & -18.0 & $5.1\times10^{-10}$ &  0.702 & 81 4f$^3$($^4$I$^\circ$)5f $^5$K  &    14 4f$^3$($^4$I$^\circ$)5f $^3$I  & \;\:2 4f$^3$($^2$H2$^\circ$)5f 3I \\
\multicolumn{1}{c}{\vdots} & \multicolumn{1}{c}{\vdots} & \multicolumn{1}{c}{\vdots} & \multicolumn{1}{c}{\vdots} & \multicolumn{1}{c}{\vdots} & \multicolumn{1}{c}{\vdots} & \multicolumn{1}{c}{\vdots} & \multicolumn{1}{c}{\vdots} & \multicolumn{1}{c}{\vdots} \\
e&                        & 8 & 130\,935.4(*)    &       & $5.2\times10^{-10}$ &  1.020 & 51 4f$^3$($^2$H2$^\circ$)5f $^1$L  &    19 4f$^3$($^2$H2$^\circ$)5f $^3$L  & 10 4f$^3$($^2$H1$^\circ$)5f $^1$L  \\
\hline
o&4f$^3$($^4$I$^\circ$)5d $^5$L$^\circ$        & 6 & 15\,158.154(7)   & -53.0 & $3.2\times10^{-3}$ &  0.724 & 93 4f$^3$($^4$I$^\circ$)5d $^5$L$^\circ$ & \;\:3 4f$^3$($^2$H2$^\circ$)5d 3K$^\circ$ & \;\:3 4f$^3$($^4$I$^\circ$)5d $^3$K$^\circ$ \\
\multicolumn{1}{c}{\vdots} & \multicolumn{1}{c}{\vdots} & \multicolumn{1}{c}{\vdots} & \multicolumn{1}{c}{\vdots} & \multicolumn{1}{c}{\vdots} & \multicolumn{1}{c}{\vdots} & \multicolumn{1}{c}{\vdots} & \multicolumn{1}{c}{\vdots} & \multicolumn{1}{c}{\vdots} \\
o&                      & 3 &  47\,499.8(*)    &       & $1.5\times10^{-6}$ &  0.926 & 53 4f$^3$($^4$G$^\circ$)6s $^5$G$^\circ$ &    17 4f$^3$($^2$G1$^\circ$)6s $^3$G$^\circ$ & 13 4f$^3$($^2$G2$^\circ$)6s $^3$G$^\circ$ \\
o&4f$^3$($^4$I$^\circ$)7s ($\frac{9}{2}$,$\frac{1}{2}$)$^\circ$ & 4 & 103\,085.765(10) &  -5.7 & $1.4\times10^{-9}$ &  0.605 & 97 4f$^3$($^4$I$^\circ$)7s $^5$I$^\circ$ & \;\:3 4f$^3$($^2$H2$^\circ$)7s $^3$H$^\circ$ & \\
o&4f$^3$($^4$I$^\circ$)7s ($\frac{9}{2}$,$\frac{1}{2}$)$^\circ$ & 5 & 103\,335.469(8)  &  -2.7 & $1.4\times10^{-9}$ &  0.866 & 56 4f$^3$($^4$I$^\circ$)7s $^3$I$^\circ$ &    40 4f$^3$($^4$I$^\circ$)7s $^5$I$^\circ$ & \;\:2 4f$^3$($^2$H2$^\circ$)7s $^3$H$^\circ$ \\
o&                       &  1 & 103\,359.1(*)    &       & $1.8\times10^{-7}$ & 1.000 & 59 4f$^3$($^2$F1$^\circ$)5d $^1$P$^\circ$ &    40 4f$^3$($^2$F2$^\circ$)5d $^1$P$^\circ$ & \\
o&4f$^3$($^4$I$^\circ$)6d $^5$K$^\circ$        & 5 & 103\,652.472(9)  &   2.6 & $1.1\times10^{-9}$ &  0.695 & 84 4f$^3$($^4$I$^\circ$)6d $^5$K$^\circ$ &    11 4f$^3$($^4$I$^\circ$)6d $^3$I$^\circ$ & \;\:3 4f$^3$($^2$H2$^\circ$)6d $^3$I$^\circ$ \\
\multicolumn{1}{c}{\vdots} & \multicolumn{1}{c}{\vdots} & \multicolumn{1}{c}{\vdots} & \multicolumn{1}{c}{\vdots} & \multicolumn{1}{c}{\vdots} & \multicolumn{1}{c}{\vdots} & \multicolumn{1}{c}{\vdots} & \multicolumn{1}{c}{\vdots} & \multicolumn{1}{c}{\vdots} \\
o&4f$^3$($^4$I$^\circ$)6d 3G$^\circ$        & 5 & 111\,138.018(18) & -26.2 & $1.3\times10^{-9}$ &  1.196 & 79 4f$^3$($^4$I$^\circ$)6d 3G$^\circ$ &    12 4f$^3$($^4$I$^\circ$)6d $^5$G$^\circ$ & \;\:4 4f$^3$($^4$I$^\circ$)6d $^3$H$^\circ$\\

%& & \vdots & \multicolumn{1}{c}{\vdots} & \multicolumn{1}{c}{\vdots} & \vdots\;\: & \vdots\;\: & \vdots & \multicolumn{1}{c}{\vdots} & \multicolumn{1}{c}{\vdots} & \multicolumn{1}{c}{\vdots}\\
%4f$^3$($^2$H2$^\circ$)6p & $^3$H & 6 & 80593.604(32) & 1(0) & 2 & -810 & 1.112 & 47 ($^2$H2$^\circ$)6p $^3$H & 21 ($^2$H2$^\circ$)6p $^1$I & \;\:9 ($^2$H1$^\circ$)6p $^3$H\\
\hline
\end{tabular}
\tablefoot{The full version of this table is available at the CDS with up to five leading eigenvector components for each energy level. The columns are: (1) parity, where `e' is for even and `o' is for odd, (2) assigned configuration and term label for levels experimentally found, (3) total angular momentum $J$, (4) experimentally determined or semi-empirically calculated level energy in this work, uncertainties are in parentheses in units of the final decimal place for experimental energies, an uncertainty `*' indicates an experimentally unknown level, where only the calculated energy is shown, (5) difference between observed and calculated level energies, (6)--(8) lifetime due to electric dipole (E1) transitions, Land\'{e} $g$-factor, and three leading eigenvector percentages calculated using the Cowan code in this work.}
\end{table*} The extract of Table \ref{tab:nd_levels} shows the lowest-lying levels of these configurations and energy ranges of the table for both parities. Typical energy uncertainties are 0.01~cm$^{-1}$ for the 4f$^3$7s and 4f$^3$6d levels and 0.3~cm$^{-1}$ for the 5f levels. We would like to note that the 4f$^3$($^4$I$^\circ$)6d $^5$H$_3$ level has relatively large energy uncertainty because the two lines observed for this level were blended with other lines in both the Nd-Ar PDL FT and the Nd VS grating spectra. The FWHM of each of the two lines in the FT spectrum was set as their statistical wavenumber uncertainties in the level energy optimisation.}

In theoretical calculations, the 4f$^3$7s levels were represented very well by the eigenfunctions from \textit{jj}-coupling the excited electron with the three 4f core electrons in the \textit{LS}-coupling scheme, similar to the 4f$^3$6s levels of Nd~III. The mean difference between all observed level energies and those from previous calculations in the literature by \cite{gaigalas2019extended} was offset within a few percent (around 6000~cm$^{-1}$), but the variance in these differences within each configuration was up to two orders of magnitude smaller than the offset. %The comparison between experimental level energies and theoretical level energies from \cite{gaigalas2019extended} was not possible for a large fraction of 4f$^3$5f levels due to the unpublished and likely incomparable eigenvector compositions in \cite{gaigalas2019extended}. %The percentage of the leading eigenvector component for many 4f$^3$5f levels calculated in the present work is less than 50\% and in some cases comparable to that of the second component, but the contribution of the second component is nearly always less than that of the first one. Term assignments were easily made using the leading component for the 4f$^3$5f levels. % except for the $J=2$ level at 115488.1~cm$^{-1}$. This level is a mixture of the 5f, 6p, and 4f$^2$5d$^2$ configurations and has the 4f$^3$($^4$I$^\circ$)5f~$^5$G term as the leading component, which was already occupied by the $J=2$ level at 115647.5~cm$^{-1}$. Therefore the 4f$^3$($^4$I$^\circ$)5f $^5$Ga assignment was given to the $J=2$ level at 115488.1~cm$^{-1}$.

%Overall, the energy level and line classifications of the 4f$^3$7s, 4f$^3$6d, and 4f$^3$5f configurations of Nd~III are more challenging in the present work compared to those of the 4f$^3$6s, and 4f$^3$6p configurations in \cite{ding2023spectrum}; the lines were lower in SNR, more dependent on the lower resolution grating spectra, and a smaller number of transitions were observed on average for each level. Nevertheless, 

\subsubsection{Classified transitions}\label{sec:lines}
In the Nd-Ar PDL FT spectra, 147 4f$^3$6p -- 4f$^3$7s and 4f$^3$6p -- 4f$^3$6d Nd~III transitions were observed, all of which were also observed in the Nd VS spectra and are listed in Table \ref{tab:nd_lines} with the maximum S/N at 37 and an average uncertainty of $\pm0.024$~cm$^{-1}$. The 39 lines of the 4f$^3$6p -- 4f$^3$7s and 4f$^3$6p -- 4f$^3$6d transition and 169 lines of the 4f$^3$5d -- 4f$^3$5f transitions observed only in the Nd VS spectra are also listed in Table \ref{tab:nd_lines}. The 39 grating lines originating from the 4f$^3$7s and 4f$^3$6d configurations were weighted zero in the level energy optimisation process; the wavelength uncertainty of these grating spectra, deduced from RMS of deviations from reference wavelengths at $\pm0.006$~\AA{}, is around $\pm0.1$~cm$^{-1}$ at 40\,000~cm$^{-1}$, which is an order of magnitude higher than the 4f$^3$7s and 4f$^3$6d level energy uncertainties optimised purely using Nd-Ar FT spectra wavenumbers. Additionally, blends are much more common in the line-rich and lower-resolution Nd VS spectra, and therefore uncertainty improvements in the optimised level energies would not be guaranteed in the inclusion of these grating lines in the level optimisation process. For the 4f$^3$5d -- 4f$^3$5f transitions observed only in the Nd VS grating spectra, their wavelength uncertainties of around $\pm0.006$~\AA{} were used in level optimisation.
{\setlength{\tabcolsep}{2pt}
\begin{sidewaystable*}
\caption{Classified transitions of Nd~III, originating from the 4f$^3$7s, 4f$^3$6d, and 4f$^3$5f configurations, observed in the Nd-Ar PDL FT and Nd VS spectra (extract).}\label{tab:nd_lines}
\centering
\footnotesize
\renewcommand{\arraystretch}{1.5}
\begin{tabular}{cccccccccccccccc}
\hline\hline             
\multicolumn{1}{c}{Spec} &
\multicolumn{1}{c}{$S/N$} & 
\multicolumn{1}{c}{FWHM} & 
\multicolumn{1}{c}{Int.} & 
\multicolumn{1}{c}{$g_uA$} & 
\multicolumn{1}{c}{log($g_lf$)} & 
\multicolumn{1}{c}{$\lambda_{\text{obs}}$} & 
\multicolumn{1}{c}{$\sigma_{\text{obs}}$} & 
\multicolumn{1}{c}{$\sigma_{\text{Ritz}}$} &
\multicolumn{1}{c}{$\sigma_{\text{obs}} - \sigma_{\text{Ritz}}$} &
\multicolumn{1}{c}{$\lambda^{\text{air}}_{\text{Ritz}}$} & 
\multicolumn{1}{c}{Lower Level} &
\multicolumn{1}{c}{Upper Level} & 
\multicolumn{1}{c}{$E_l$} & 
\multicolumn{1}{c}{$E_u$} &
\multicolumn{1}{c}{Note}
\\
&
&
\multicolumn{1}{c}{(cm$^{-1}$)} &
\multicolumn{1}{c}{(arb.)} & 
\multicolumn{1}{c}{(s$^{-1}$)} & 
&
\multicolumn{1}{c}{(\AA)} &
\multicolumn{1}{c}{(cm$^{-1}$)} & 
\multicolumn{1}{c}{(cm$^{-1}$)} & 
\multicolumn{1}{c}{(cm$^{-1}$)} & 
\multicolumn{1}{c}{(\AA)} & 
\multicolumn{1}{c}{Label} &% Term$_J$ &
\multicolumn{1}{c}{Label} &% Term$_J$ &
\multicolumn{1}{c}{(cm$^{-1}$)} & 
\multicolumn{1}{c}{(cm$^{-1}$)} & 
\\
\multicolumn{1}{c}{(1)} & 
\multicolumn{1}{c}{(2)} & 
\multicolumn{1}{c}{(3)} & 
\multicolumn{1}{c}{(4)} & 
\multicolumn{1}{c}{(5)} & 
\multicolumn{1}{c}{(6)} & 
\multicolumn{1}{c}{(7)} & 
\multicolumn{1}{c}{(8)} & 
\multicolumn{1}{c}{(9)} & 
\multicolumn{1}{c}{(10)} & 
\multicolumn{1}{c}{(11)} & 
\multicolumn{1}{c}{(12)} & 
\multicolumn{1}{c}{(13)} & 
\multicolumn{1}{c}{(14)} & 
\multicolumn{1}{c}{(15)} &
\multicolumn{1}{c}{(16)}\\
\hline
g &    &       &  1 & 3.1$\times10^9$ & -0.34 &  993.794(6)   & 100\,624.4(6) & 100\,624.528(300) & -0.128 & 993.793(3) & 4f$^3$($^4$I$^\circ$)5d $^5$K$^\circ$$_7$ & 4f$^3$($^4$I$^\circ$)5f $^5$I$_6$ & 18\,656.272 & 119\,280.8 &  \\
g &    &       &  1 & 2.2$\times10^9$ & -0.49 &  994.019(6)   & 100\,601.7(6) & 100\,601.732(300) &  -0.032 & 994.019(3) & 4f$^3$($^4$I$^\circ$)5d $^5$K$^\circ$$_6$ & 4f$^3$($^4$I$^\circ$)5f $^5$I$_5$ & 16\,938.068 & 117\,539.8 &  \\
g &    &       &  0 & 1.8$\times10^9$ & -0.57 &  994.138(6)   & 100\,589.7(6) & 100\,589.863(300) & -0.163 & 994.136(3) & 4f$^3$($^4$I$^\circ$)5d $^5$K$^\circ$$_5$ & 4f$^3$($^4$I$^\circ$)5f $^5$I$_4$ & 15\,262.437 & 115\,852.3 &  \\
g &    &       &  3 & 2.4$\times10^9$ & -0.45 &  997.327(6)   & 100\,268.0(6) & 100\,268.628(300) & -0.628 & 997.321(3) & 4f$^3$($^4$I$^\circ$)5d $^5$K$^\circ$$_7$ & 4f$^3$($^4$I$^\circ$)5f $^3$L$_8$ & 18\,656.272 & 118\,924.9 &  \\
g &    &       &  5 & 4.8$\times10^9$ & -0.14 &  997.916(6)   & 100\,208.8(6) & 100\,208.936(400) & -0.136 & 997.915(4) & 4f$^3$($^4$I$^\circ$)5d $^5$L$^\circ$$_8$ & 4f$^3$($^4$I$^\circ$)5f $^3$M$_9$ & 18\,861.064 & 119\,070.0 &  \\
\multicolumn{1}{c}{\vdots}&\multicolumn{1}{c}{\vdots}&\multicolumn{1}{c}{\vdots}&\multicolumn{1}{c}{\vdots}&\multicolumn{1}{c}{\vdots}&\multicolumn{1}{c}{\vdots}&\multicolumn{1}{c}{\vdots}&\multicolumn{1}{c}{\vdots}&\multicolumn{1}{c}{\vdots}&\multicolumn{1}{c}{\vdots}&\multicolumn{1}{c}{\vdots}&\multicolumn{1}{c}{\vdots}&\multicolumn{1}{c}{\vdots}&\multicolumn{1}{c}{\vdots}&\multicolumn{1}{c}{\vdots}&\multicolumn{1}{c}{\vdots} \\
g &  &  & 5 & 8.1$\times10^9$ & 0.15 & 1083.318(6) & 92\,309.0(5) & 92309.079(400) & -0.079 & 1083.317(5) & 4f$^3$($^4$I$^\circ$)5d $^3$L$^\circ$$_9$ & 4f$^3$($^4$I$^\circ$)5f $^3$L$_9$ & 28\,877.321 & 121\,186.4 &  \\
g &  &  & 30 & 6.4$\times10^9$ & 0.05 & 1084.030(6) & 92\,248.4(5) & 92248.215(300) & 0.185 & 1084.032(4) & 4f$^3$($^4$I$^\circ$)5d $^3$L$^\circ$$_8$ & 4f$^3$($^4$I$^\circ$)5f $^3$L$_8$ & 26\,676.685 & 118\,924.9 &  \\
g &  &  & 40 & 2.2$\times10^{10}$ & 0.59 & 1084.129(6) & 92\,239.9(5) & 92239.935(400) & -0.035 & 1084.129(5) & 4f$^3$($^4$I$^\circ$)5d $^3$L$^\circ$$_7$ & 4f$^3$($^4$I$^\circ$)5f $^3$M$_8$ & 24\,497.165 & 116\,737.1 &  \\
g &    &       &  2 & 3.9$\times10^9$ & -0.16 & 1085.156(6)   & 92\,152.6(5) & 92\,152.735(300) & -0.135 & 1085.155(4) & 4f$^3$($^4$I$^\circ$)5d $^3$L$^\circ$$_7$ & 4f$^3$($^4$I$^\circ$)5f $^3$L$_7$ & 24\,497.165 & 116\,649.9 &  \\
g &    &       &  8 & 6.3$\times10^9$ &  0.06 & 1107.003(6)   & 90\,334.0(5) & 90\,334.035(300) & -0.035 & 1107.002(4) & 4f$^3$($^4$I$^\circ$)5d $^3$L$^\circ$$_7$ & 4f$^3$($^4$I$^\circ$)5f $^5$M$_8$ & 24\,497.165 & 114831.2 &  \\
\hline
g &    &       & 19 & 3.1$\times10^7$ & -1.66 & 2164.085(6)   & 46\,208.92(13) & 46\,209.035(13) & -0.118 & 2163.3997(6) & 4f$^3$($^4$I$^\circ$)6p $^5$K$_6$ & 4f$^3$($^4$I$^\circ$)6d $^3$K$^\circ$$_7$ & 62\,520.646 & 108\,729.681 &  \\
f &  4 & 0.210 & 22 & 4.4$\times10^9$ &  0.53 & 2254.5500(17) & 44\,354.749(33) & 44\,354.728(9) & 0.021 & 2253.8528(5) & 4f$^3$($^4$I$^\circ$)6p $^5$I$_8$ & 4f$^3$($^4$I$^\circ$)6d $^3$L$^\circ$$_9$ & 66\,792.024 & 111\,146.752 &  \\
f & 6 & 0.144 & 24 & 3.3$\times10^9$ & 0.40 & 2260.6456(9) & 44\,235.151(17) & 44\,235.150(10) & 0.001 & 2259.9461(5) & 4f$^3$($^4$I$^\circ$)6p $^5$K$_7$ & 4f$^3$($^4$I$^\circ$)6d $^3$L$^\circ$$_8$ & 64\,622.006 & 108\,857.156 &  \\
g &  &  & 21 & 3.7$\times10^8$ & -0.55 & 2263.457(6) & 44\,180.21(12) & 44\,180.061(12) & 0.150 & 2262.7643(6) & 4f$^3$($^4$I$^\circ$)6p $^5$I$_6$ & 4f$^3$($^4$I$^\circ$)6d $^3$K$^\circ$$_7$ & 64\,549.620 & 108\,729.681 &  \\
g &  &  & 49 & 7.1$\times10^8$ & -0.26 & 2264.787(6) & 44\,154.26(12) & 44\,154.232(14) & 0.031 & 2264.0881(7) & 4f$^3$($^4$I$^\circ$)6p $^5$I$_4$ & 4f$^3$($^4$I$^\circ$)6d $^5$G$^\circ$$_3$ & 60\,638.963 & 104\,793.195 &  \\
\multicolumn{1}{c}{\vdots}&\multicolumn{1}{c}{\vdots}&\multicolumn{1}{c}{\vdots}&\multicolumn{1}{c}{\vdots}&\multicolumn{1}{c}{\vdots}&\multicolumn{1}{c}{\vdots}&\multicolumn{1}{c}{\vdots}&\multicolumn{1}{c}{\vdots}&\multicolumn{1}{c}{\vdots}&\multicolumn{1}{c}{\vdots}&\multicolumn{1}{c}{\vdots}&\multicolumn{1}{c}{\vdots}&\multicolumn{1}{c}{\vdots}&\multicolumn{1}{c}{\vdots}&\multicolumn{1}{c}{\vdots}&\multicolumn{1}{c}{\vdots} \\
f &  7 & 0.239 & 24 & 1.4$\times10^9$ &  0.15 & 2599.5239(13) & 38\,468.583(19) & 38\,468.575(11) & 0.008 & 2598.7476(7) & 4f$^3$($^4$I$^\circ$)6p $^3$K$_8$ & 4f$^3$($^4$I$^\circ$)7s ($\frac{15}{2}$,$\frac{1}{2}$)$^\circ$$_8$ & 70\,599.570 & 109\,068.145 &  \\
f &  6 & 0.283 & 26 & 7.9$\times10^8$ & -0.10 & 2603.2166(16) & 38\,414.015(24) & 38\,414.008(11) & 0.007 & 2602.4394(7) & 4f$^3$($^4$I$^\circ$)6p $^5$H$_7$ & 4f$^3$($^4$I$^\circ$)7s ($\frac{11}{2}$,$\frac{1}{2}$)$^\circ$$_6$ & 66\,717.513 & 105\,131.521 &  \\
f & 18 & 0.214 & 62 & 2.2$\times10^9$ &  0.35 & 2610.1592(5)  & 38\,311.840(7) & 38\,311.836(6) & 0.004 & 2609.3801(4) & 4f$^3$($^4$I$^\circ$)6p $^5$H$_6$ & 4f$^3$($^4$I$^\circ$)7s ($\frac{9}{2}$,$\frac{1}{2}$)$^\circ$$_5$ & 65\,023.633 & 103\,335.469 &  \\
f &  2 & 0.170 &  6 & 6.9$\times10^7$ & -1.13 & 2679.0382(37) & 37\,326.829(52) & 37\,326.835(12) & -0.006 & 2678.2420(9) & 4f$^3$($^4$I$^\circ$)6p $^3$K$_8$ & 4f$^3$($^4$I$^\circ$)6d $^5$K$^\circ$$_8$ & 70\,599.570 & 107\,926.405 &  \\
g &    &       &  8 & 2.0$\times10^8$ & -0.66 & 2711.225(6)   & 36\,883.69(8) & 36\,883.657(8) & 0.035 & 2710.4244(6) & 4f$^3$($^4$I$^\circ$)6p $^3$K$_6$ & 4f$^3$($^4$I$^\circ$)7s ($\frac{9}{2}$,$\frac{1}{2}$)$^\circ$$_5$ & 66\,451.812 & 103\,335.469 &  \\
\hline
\end{tabular}
\tablefoot{The full electronic version of this table is available at the CDS. The top half of this extract shows observed 4f$^3$5d -- 4f$^3$5f transitions in the grating spectra, the bottom half shows 4f$^3$6p -- 4f$^3$(7s + 6d) transitions observed in both the grating and FT spectra. The columns are: (1) spectrum of observation, where `g' indicates line observed only in the grating Nd VS spectra and `f' indicates line observed in both the Nd-Ar PDL FT and Nd VS spectra, where only the higher accuracy FT spectral lines are presented, (2)--(3) signal-to-noise ratio and full width at half maximum of the fitted FT spectral line, (4) approximate relative intensity corresponding to relative photon flux for `f' and to relative energy flux for `g', where lines above 2000~\AA{} are on the same scale as those in Tables 6 and 7 of \cite{ding2023spectrum} and lines below 2000~\AA{} scale between 0 and 100. (5)--(6) weighted TP and log of the weighted (absorption) oscillator strength calculated using the Cowan code, where $g_u$ and $g_l$ refer to statistical weights of the upper and lower energy levels, respectively, (7)--(8) observed vacuum wavelength for `g' and vacuum wavenumber for `f', (9) Ritz wavenumber from level optimisation, (10) wavenumber difference between observed and Ritz values, (11) Ritz air wavelength for lines above 2000~\AA{} converted using the three-term 
dispersion formula from \cite{peck1972dispersion}, Ritz vacuum wavelengths are given for lines below 2000~\AA{}, (12)--(13) energy levels associated with the transition, their energies are in columns (14)--(15), respectively, and (16) contains comments of the observed transition, where `B/W' indicates a blended or weak line with unreliable wavenumber and intensity, which was omitted from level optimisation. Uncertainties of columns (7), (8), and (9) are in parentheses in units of the final decimal place.}
\end{sidewaystable*}}

We would like to note that the relative intensity measurements in Table \ref{tab:nd_lines} are recommended only as guides \citep{ding2023spectrum}. For the FT spectral lines, we estimate relative intensity uncertainties of at least 20\% due to low S/Ns. These are also expected to be larger when comparing lines between FT spectra E and F, primarily due to the different lamp conditions on different days. For the grating spectral lines, the uncertainties are expected to be even more uncertain, and relative intensities are recommended to be qualitative as these are the photographic plate darkening calibrated using approximate sensitivity and grating instrumental response curves.

\subsection{Analysis and remarks}\label{analysis_remarks}
%For brevity, the 4f$^3$ core electrons in the rest of this paper will be omitted in any references to the singly-excited configurations, e.g., levels of the 4f$^3$6p configuration will be referred to as 6p levels, full labels will still be used when referring to specific sub-configurations and energy levels.
We discuss the analysis and finding of the new energy levels of this work in this section. We expect most energy levels to be correctly identified and that they will improve current constraints and benchmarks for large-scale Nd~III atomic structure calculations. For the 4f$^3$6d levels with only one observed transition in Table \ref{tab:nd_lines}, caution is advised and they should be more appropriately treated as tentative. For each of the 4f$^3$6d levels of this category, the strongest transition to the 4f$^3$6p configuration was observed across both the Nd-Ar PDL FT and Nd VS grating spectra and was the only candidate within $\sim$50~cm$^{-1}$ of the predicted wavenumber.

\subsubsection{\texorpdfstring{Finding 4f$^3$($^2$H2$^\circ$)5d $^3$F$^\circ$$_4$}{} in stellar spectra}
\begin{figure*}[!t]
     \includegraphics[width=\linewidth]{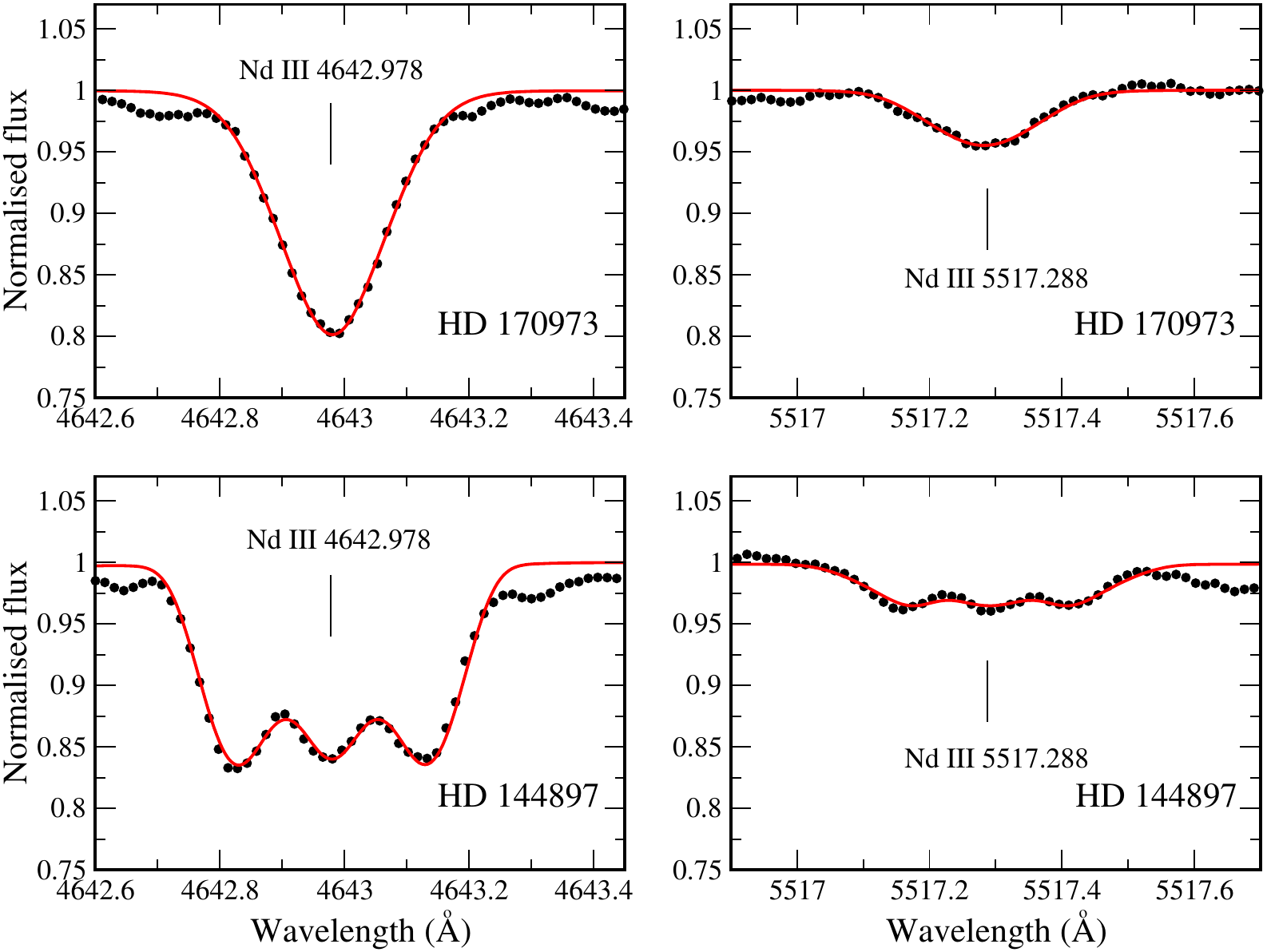}
     \caption{Observed (black filled circles) and sythesised (red line) spectral lines of the newly identified 4f$^3$($^2$H2$^\circ$)5d $^3$F$^\circ$$_4$ level in spectra of HD~170973 and HD~144893 with surface magnetic fields \bs$<$1~kG and \bs=8.8~kG, respectively, showing two newly classified, previously unknown transitions of Nd~III: 4f$^4$ $^5$F$_5$ -- 4f$^3$($^2$H2$^\circ$)5d $^3$F$^\circ$$_4$ within 4642.6--4643.4~\AA{} and 4f$^4$ $^3$H$_4$ -- 4f$^3$($^2$H2$^\circ$)5d $^3$F$^\circ$$_4$ within 5516.9--5517.7~\AA{}.}
     \label{fig1}
 \end{figure*}

One additional level of the previously known 4f$^3$5d configuration is newly identified based on the fitting of calculated line intensities and Zeeman patterns to the absorption line profiles in stellar spectra described in \cite{ding2023spectrum}. It is given the label 4f$^3$($^2$H2$^\circ$)5d $^3$F$^\circ$$_4$ with energy estimated at $34\,742.15\pm0.10$~cm$^{-1}$. Two transitions from 4f$^3$($^2$H2$^\circ$)5d $^3$F$^\circ$$_4$ to known levels 4f$^4$ $^5$F$_5$ (4642.97~\AA{}, the stronger line) and 4f$^4$ $^3$H$_4$ (5517.27~\AA{}, the weaker line) were found in stellar spectra, their fits are shown in Fig. \ref{fig1}.
 The stronger transition is the only transition observed in the Nd-Ar FT spectra, but appeared blended in both the PDL and  HCL spectra. 

\subsubsection{Initial Cowan code parameters}
Before the experimental investigation of the 4f$^3$7s, 4f$^3$6d, and 4f$^3$5f levels, theoretical atomic structure calculations for these configurations were made using the Cowan codes \citep{cowan1981theory, kramida2021suite}, as these would guide the process of finding levels. The scaling factors of the 4f$^3$6d configurations in these Cowan code calculations were taken as those fitted for the 4f$^3$5d configuration. Parameters describing the interactions within the 4f$^3$ sub-shell in the 4f$^3$7s and 4f$^3$5f configurations were also taken from their empirical ratios determined for the 5d configuration. Lastly, the 4f$^4$ -- 4f$^3$7s and 4f$^4$ -- 4f$^3$5f interactions were scaled by 0.85 with respect to their corresponding Hartree-Fock values. The Cowan code calculations used for identifying levels of the present work were presented in \cite{ding2023spectrum}.

\subsubsection{\texorpdfstring{4f$^3$($^4$I$^\circ$)7s and 4f$^3$($^4$I$^\circ$)6d levels}{4f3(4I)7s and 4f3(4I)6d levels}}
All 8 levels of the 4f$^3$($^4$I$^\circ$)7s sub-configuration were identified and 38 of the 40 levels of the 4f$^3$($^4$I$^\circ$)6d sub-configuration were identified. The 4f$^3$6d -- 4f$^3$5f lines are around 8000~\AA{} and calculated to have high TPs, but they lie outside the available sensitive spectral range of the Imperial College VUV FT spectrometer, so the known 4f$^3$5f levels could not be used in the search for the unknown 4f$^3$6d levels or vice versa. 

As the 4f$^3$6p configuration of Nd~III is only well established for the 4f$^3$($^4$I$^\circ$) parent term, only new 4f$^3$($^4$I$^\circ$)7s and 4f$^3$($^4$I$^\circ$)6d levels were identified in the present work, through classification of the newly observed 4f$^3$6p -- 4f$^3$7s and 4f$^3$6p -- 4f$^3$6d transitions. Attempts for these classifications were not successful using only the Nd-Ar PDL FT spectra and theoretical calculations from \cite{gaigalas2019extended}, where the 4f$^3$7s and 4f$^3$6d levels were predicted to have energies around 6000~cm$^{-1}$ higher than their eventually observed values. Confident classifications of the transitions were first made in the Nd VS spectra, followed by confirmations with corresponding weak but more wavelength-accurate lines in the Nd-Ar PDL FT spectra. This initial progress improved the Cowan code parameter constraints and provided expectations for similar relative intensities in the FT and grating spectra for the other 4f$^3$6p -- 4f$^3$7s and 4f$^3$6p -- 4f$^3$6d transitions.

\begin{figure}
    \centering
    \includegraphics[width=\linewidth]{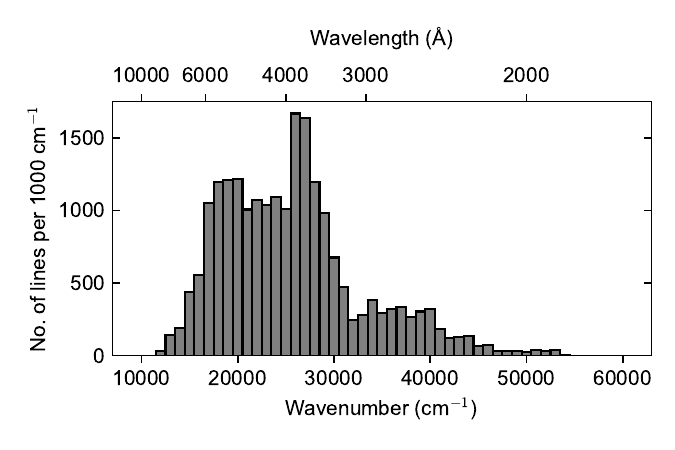}
    \caption{The number of lines per 1000~cm$^{-1}$ observed in the six Nd-Ar PDL FT spectra. We note that this distribution is largely dependent on the experimental parameters \citep[see Table 2 of][]{ding2023spectrum}, e.g., instrumental response and discharge conditions.}
    \label{fig2}
\end{figure}

The two UV Nd-Ar PDL FT spectra contain only 3451 of the 21\,584 lines fitted in total from all six spectra measured across the spectral range 11\,500--54\,000~cm$^{-1}$ (8695--1852~\AA{}) in \cite{ding2023spectrum}. The line densities across the spectral range are illustrated in Fig. \ref{fig2}. 
Compared to classifying the 4f$^4$ -- 4f$^3$5d transitions in the visible region, the lower line density and fewer number of blended lines above $\sim$30\,000~cm$^{-1}$ produced very few ambiguous classifications for the 4f$^3$5d -- 4f$^3$6p, 4f$^3$6s -- 4f$^3$6p, 4f$^3$6p -- 4f$^3$7s, and 4f$^3$6p -- 4f$^3$6d transitions predicted \citep{gaigalas2019extended, ding2023spectrum} to lie within this region. While the 4f$^3$5d -- 4f$^3$6p and 4f$^3$6s -- 4f$^3$6p transitions were relatively straightforwardly classified in \cite{ding2023spectrum}, this was far more challenging for many 4f$^3$6p -- 4f$^3$7s and 4f$^3$6p -- 4f$^3$6d transitions; the strongest lines of these four transition arrays have similar predicted TPs, but the 7s and 6d levels lie about 40\,000~cm$^{-1}$ higher than the 6p levels. Therefore, due to lower level populations of the 4f$^3$7s and 4f$^3$6d levels, fewer 4f$^3$6p -- 4f$^3$7s and 4f$^3$6p -- 4f$^3$6d transitions were observed and were observed at lower S/Ns compared to the 4f$^3$5d -- 4f$^3$6p and 4f$^3$6s -- 4f$^3$6p transitions in the Nd-Ar PDL FT spectra. The reduction in line S/Ns and in the number of lines observed from each energy level increased the number of ambiguous classifications, hence many 4f$^3$7s and 4f$^3$6d levels could not be identified without the Nd VS grating spectra containing the weaker lines, especially in cases when only one line of a 4f$^3$6d level was expected to be observable in the Nd-Ar PDL FT spectra.

In contrast, the number of lines measured in the Nd VS grating spectra between 1921-3250~\AA{} was around 9000 and these were recorded under spectral resolving powers around one order of magnitude lower than those in the Nd-Ar FT spectra, where blended lines were expected to be very common at this line density. However, due to the much higher discharge currents, many 4f$^3$6p -- 4f$^3$7s and 4f$^3$6p -- 4f$^3$6d transitions were observed and their levels were confidently identified solely using the Nd VS grating spectra, which were later found to be consistent with the lines observed near the noise level in the Nd-Ar PDL FT spectra. Nevertheless, it was not possible to find the 4f$^3$7s and 4f$^3$6d levels with only a few transitions to known levels without corroboration with the Nd-Ar PDL FT spectra, e.g., many incorrect candidate lines were not observed or observed at unlikely relative intensities in the Nd-Ar PDL FT spectra. 

\subsubsection{\texorpdfstring{4f$^3$5f levels}{4f35f levels}}
Based on the classification of 169 lines of the 4f$^3$5d -- 4f$^3$5f transitions, 70 levels of the 5f configuration were found, containing all 56 levels of the 4f$^3$($^4$I$^\circ$)5f sub-configuration and 14 levels of the 4f$^3$($^4$F$^\circ$)5f and 4f$^3$($^2$H2$^\circ$)5f sub-configurations. Transitions involving the 4f$^3$5f levels were not observed in the Nd-Ar PDL FT spectra. As was mentioned previously, the wavelengths and intensities of the 4f$^3$5d -- 4f$^3$5f transitions were taken from the line list of \cite{wyart2006private} in the range 820--1159~\AA{}. Only lines with intensities greater than 10 on this line list scale were definitely present in the Troitsk spectra. However, the intensity differences between the line lists of \cite{wyart2006private} and Troitsk (of spectra recorded at two different spark currents) were used to aid classifications of lines belonging to Nd~III.

\section{New Nd III atomic structure and transition probability calculations}
As the experimental energy level and spectrum analysis of Nd~III concluded, final semi-empirical Cowan code calculations of Nd~III fine structure and TPs were carried out. Results from these calculations are presented in this section. 
\subsection{Cowan Code parameters}
The previous analysis of Nd~III \citep{ding2023spectrum} did not include configurations involving the excitation of a 5p electron of the 5p$^6$ core, where underestimations of the experimental lifetimes of five levels of Nd~III \citep{zhang2002measurement} were evident. Configuration interactions involving these 5p-excited configurations were shown to significantly increase calculated level lifetimes of Nd IV \citep{wyart2007analysis,arab2019observation} and Nd~V \citep{meftah2008spectrum,deghiche2015observation}. 
To account for this in Nd~III and to check the consistency of our energy level identifications, new semi-empirical calculations for Nd~III were carried out using the Cowan code \citep[e.g.][]{cowan1981theory,kramida2021suite} in the present work. The calculations presented here were made after the identification of all Nd~III energy levels of this work.

Including all known configurations of Nd~III and significant 5p-excited configurations of Nd~III led to prohibitively large Hamiltonian matrix dimensions for our computational resources. Therefore, we included only the most significant configuration interactions of the 5p-excited configurations with the previously known and newly established energy levels, these are 5p$^6$4f$^4$ $\leftrightarrow$ 5p$^5$4f$^5$ and 5p$^6$4f$^3$6p $\leftrightarrow$ 5p$^5$4f$^4$6p for the even parity, and 5p$^6$4f$^3$5d $\leftrightarrow$ 5p$^5$4f$^4$5d, 5p$^6$4f$^3$6s $\leftrightarrow$ 5p$^5$4f$^4$6s, 5p$^6$4f$^3$7s $\leftrightarrow$ 5p$^5$4f$^4$7s, and 5p$^6$4f$^3$6d $\leftrightarrow$ 5p$^5$4f$^4$6d for the odd parity. Separate configuration interaction spaces were used in calculations for the 4f$^3$(4f, 5f, 5d, 6s, and 6p) configurations (case 1) and 4f$^3$(7s + 6d) configurations (case 2). These two cases are presented in Table \ref{tab:CI_space}, 
\begin{table}[]
    \caption{Configuration interaction space for the 4f$^3$(4f, 5f, 5d, 6d, and 6p) (case 1) and 4f$^3$(7s + 6d) (case 2) singly excited configurations of Nd~III.}
    \label{tab:CI_space}
    \centering
    \renewcommand{\arraystretch}{1.3}
    \begin{tabular}{lc}
    \hline\hline
         Parity & Configurations \\\hline
         Even (both cases) &  (4f$^4$, 5p$^5$4f$^5$), (4f$^3$6p, 5p$^5$4f$^4$6p), \\ & 4f$^2$5d$^2$, 4f$^3$5f, 4f$^2$5d6s.\\\hline
         Odd (case 1)& (4f$^3$5d, 5p$^5$4f$^4$5d), 4f$^3$6d, \\&(4f$^3$6s, 5p$^5$4f$^4$6s), 4f$^3$7s, 4f$^2$5d6p.  \\
         Odd (case 2)&  4f$^3$5d, (4f$^3$6d, 5p$^5$4f$^4$6d), 4f$^3$6s, \\ & (4f$^3$7s, 5p$^5$4f$^4$7s), 4f$^2$5d6p. \\\hline
    \end{tabular}
\end{table}
where the aforementioned significant configuration interactions are indicated in parentheses.
%In both cases, the even configuration interaction space was 4f$^4$, 4f$^3$6p, 4f$^2$5d$^2$, 4f$^2$5d6s, 5p$^5$4f$^5$, and 5p$^5$4f$^4$6p. However, the odd configurations included for the calculations of the 4f$^3$(4f + 5d + 6d + 6p) levels were 4f$^3$5d, 4f$^3$6d, 4f$^3$6s, 4f$^2$5d6p, 5p$^5$4f$^4$5d, and 5p$^5$4f$^4$6s, those for the 4f$^3$(7s + 6d) levels were 4f$^3$5d, 4f$^3$6d, 4f$^3$7s, 4f$^2$5d6p, 5p$^5$4f$^4$6d, and 5p$^5$4f$^4$7s. 
%Hence, the energy level and TP calculations used to find levels of the 4f$^3$5f configuration are those from \cite{ding2023spectrum}.

Cowan code energy parameters for the 4f$^3$(7s, 6d, and 5f) configurations are listed in Table \ref{tab:cowan_params}. 
\begin{table}[!ht]
\footnotesize
\renewcommand{\arraystretch}{1.3}
\caption{Parameters of the least-squares fit of energy levels of the 4f$^3$7s, 4f$^3$6d, and 4f$^3$5f configurations of Nd~III in the Cowan codes (extract). \label{tab:cowan_params}}
\centering
\begin{tabular}{rlrcrr}
\hline\hline
Conf. &
Param. &
LSF\tablefootmark{a} &
G\tablefootmark{b} &
HFR\tablefootmark{a} &
Ratio\tablefootmark{a,c}\\
 &
 &
(cm$^{-1}$) &
 &
(cm$^{-1}$) &
 \\
 \hline
4f$^3$7s& $E_{\text{av}}$ & 128\,608(13) &        & 120\,146        & 8462       \\
        & $F^2$(4f,4f)    & 78\,159(f)   &        & 102\,509       & 0.762      \\
        & $F^4$(4f,4f)    & 53\,201(f)   &        & 64\,317        & 0.827      \\
        & $F^6$(4f,4f)    & 36\,574(f)   &        & 46\,270        & 0.790      \\
        & $\alpha$(4f)    & 22(f)      &        & 0            &            \\
        & $\beta$(4f)     & -600(f)    &        & 0            &            \\
        & $\gamma$(4f)    & 1450(f)    &        & 0            &            \\
        & $\zeta$(4f)     & 891(2)     & 1      & 956          & 0.932      \\
        & $G^3$(4f,7s)    & 714(76)    &        & 917          & 0.779      \\
\hline
%Odd     &                 &            &        &              &            \\
4f$^3$6d& $E_{\text{av}}$ & 129\,864(6)  &        & 121\,138        & 8726       \\
        & $F^2$(4f,4f)    & 78\,094(f)   &        & 102\,424       & 0.762      \\
        & $F^4$(4f,4f)    & 53\,153(f)   &        & 64\,259        & 0.827      \\
        & $F^6$(4f,4f)    & 36\,540(f)   &        & 46\,227        & 0.790      \\
        & $\alpha$(4f)    & 22(f)      &        & 0            &            \\
        & $\beta$(4f)     & -600(f)    &        & 0            &            \\
        & $\gamma$(4f)    & 1450(f)    &        & 0            &            \\
        & $\zeta$(4f)     & 890(2)     & 1      & 955          & 0.932      \\
        & $\zeta$(6d)     & 200(7)     &        & 174          & 1.146      \\
%        & $F^1$(4f,6d)    & 0(f)       &        & 0            &            \\
        & $F^2$(4f,6d)    & 4389(183)  &        & 5283        & 0.831      \\
%        & $F^3$(4f,6d)    & 0(f)       &        & 0            &            \\
        & $F^4$(4f,6d)    & 2251(338)  &        & 2328        & 0.967      \\
        & $G^1$(4f,6d)    & 1372(56)   &        & 1938        & 0.708      \\
%        & $G^2$(4f,6d)    & 0(f)       &        & 0            &            \\
        & $G^3$(4f,6d)    & 1782(203)  &        & 1723        & 1.034      \\
%        & $G^4$(4f,6d)    & 0(f)       &        & 0            &            \\
        & $G^5$(4f,6d)    & 1172(115)  &        & 1355         & 0.864      \\
\hline
4f$^3$5f& $E_{\text{av}}$ & 139\,658(21) &        & 131\,307        & 8351      \\
        & $F^2$(4f,4f)    & 76\,479(124) &        & 102\,542       & 0.746      \\
        & $F^4$(4f,4f)    & 52\,758(167) &        & 64\,340        & 0.820      \\
        & $F^6$(4f,4f)    & 36\,567(f)   &        & 46\,287        & 0.790      \\
        & $\alpha$(4f)    & 22(f)      &        & 0            &            \\
        & $\beta$(4f)     & -600(f)    &        & 0            &            \\
        & $\gamma$(4f)    & 1450(f)    &        & 0            &            \\
        & $\zeta$(4f)     & 885(2)     &        & 956          & 0.926      \\
        & $\zeta$(5f)     &  21(7)     &        & 21          & 1.000      \\
%        & $F^1$(4f,5f)    & 0(f)       &        & 0            &            \\
        & $F^2$(4f,5f)    & 2579(131)  &        & 4027        & 0.641      \\
%        & $F^3$(4f,5f)    & 0(f)       &        & 0            &            \\
        & $F^4$(4f,5f)    & 1404(282)  &        & 1575        & 0.891      \\
%        & $F^5$(4f,5f)    & 0(f)       &        & 0            &            \\
        & $F^6$(4f,5f)    & 877(148)   &        & 1010        & 0.868      \\
        & $G^0$(4f,5f)    & 879(16)    &        & 2811        & 0.313      \\
%        & $G^1$(4f,5f)    & 0(f)       &        & 0            &            \\
        & $G^2$(4f,5f)    & 990(118)   &        & 1846        & 0.536      \\
%        & $G^3$(4f,5f)    & 0(f)       &        & 0            &            \\
        & $G^4$(4f,5f)    & 976(241)   &        & 1238         & 0.788      \\
%        & $G^5$(4f,5f)    & 0(f)       &        & 0            &            \\
        & $G^6$(4f,5f)    & 788(105)   &        & 911         & 0.864      \\
\hline
\end{tabular}
\tablefoot{The full version of this table is available at the CDS, including parameters corresponding to the interaction space of Table \ref{tab:cowan_params}.
\tablefoottext{a}{Parameter values determined in the ab initio pseudo-relativistic Hartree–Fock (HFR) and least-squares-fitted (LSF) calculations and their ratio. Standard deviations of the fitted LSF parameters are in parentheses, where `f' means the parameter was fixed.}
\tablefoottext{b}{Parameters in each numbered group (G) were linked together by sharing the same ratios to their corresponding HFR values.}
\tablefoottext{c}{Differences between LSF and HFR parameters are given for $E_{\text{av}}$ in this extract.}
}
%\vspace{0.5cm} % to make table the sole occupier of the column
\end{table}
Configuration interaction parameters, and parameters of other configurations of Table \ref{tab:CI_space} are available in the full version of Table \ref{tab:cowan_params} at the CDS. All Slater and configuration interaction parameters of the 5p-excited configurations were fixed at 0.85 and 0.7 of their ab initio HFR values, respectively. The calculated energy levels are presented in Table \ref{tab:nd_levels}, but levels predicted in the odd system with energies between 47\,500 and 103\,085~cm$^{-1}$ are not included, because they are expected to be very uncertain due to the lack of known levels in this energy range in the odd parity.
\subsection{Transition probabilities and energy level lifetimes}
The calculated TPs are presented in Table \ref{tab:nd_tps}.{\setlength{\tabcolsep}{4pt}
\begin{sidewaystable*}
\caption{Calculated TPs using configuration interaction spaces of Table \ref{tab:CI_space} (extract).}\label{tab:nd_tps}
\footnotesize
\centering
\renewcommand{\arraystretch}{1.5}
\begin{tabular}{ccccccccccccccc}
\hline\hline             
\multicolumn{1}{c}{Type} &
\multicolumn{1}{c}{CF} & 
\multicolumn{1}{c}{$g_uA$} & 
\multicolumn{1}{c}{log($g_lf$)} & 
\multicolumn{1}{c}{$\sigma_{\text{Ritz}}$} &
\multicolumn{1}{c}{$\lambda_{\text{Ritz}}$} & 
\multicolumn{1}{c}{$\lambda^{\text{air}}_{\text{Ritz}}$} & 
\multicolumn{1}{c}{Lower Level} &
\multicolumn{1}{c}{$J_l$} &
\multicolumn{1}{c}{Upper Level} & 
\multicolumn{1}{c}{$J_u$} &
\multicolumn{1}{c}{$E_l$} & 
\multicolumn{1}{c}{$E_u$} &
\multicolumn{1}{c}{$g^l_l$} & 
\multicolumn{1}{c}{$g^l_u$} 
%\multicolumn{1}{c}{$\tau_l$} &
%\multicolumn{1}{c}{$\tau_u$}
\\
\multicolumn{1}{c}{} &
\multicolumn{1}{c}{} & 
\multicolumn{1}{c}{(s$^{-1}$)} & 
\multicolumn{1}{c}{} & 
\multicolumn{1}{c}{(cm$^{-1}$)} &
\multicolumn{1}{c}{(\AA)} & 
\multicolumn{1}{c}{(\AA)} & 
\multicolumn{1}{c}{Term Label} &
\multicolumn{1}{c}{} &
\multicolumn{1}{c}{Term Label} & 
\multicolumn{1}{c}{} &
\multicolumn{1}{c}{(cm$^{-1}$)} & 
\multicolumn{1}{c}{(cm$^{-1}$)} &
\multicolumn{1}{c}{} & 
\multicolumn{1}{c}{} 
%\multicolumn{1}{c}{(ns)} &
%\multicolumn{1}{c}{(ns)}
\\
\multicolumn{1}{c}{(1)} & 
\multicolumn{1}{c}{(2)} & 
\multicolumn{1}{c}{(3)} & 
\multicolumn{1}{c}{(4)} & 
\multicolumn{1}{c}{(5)} & 
\multicolumn{1}{c}{(6)} & 
\multicolumn{1}{c}{(7)} & 
\multicolumn{1}{c}{(8)} & 
\multicolumn{1}{c}{(9)} & 
\multicolumn{1}{c}{(10)} & 
\multicolumn{1}{c}{(11)} & 
\multicolumn{1}{c}{(12)} & 
\multicolumn{1}{c}{(13)} & 
\multicolumn{1}{c}{(14)} & 
\multicolumn{1}{c}{(15)} \\ 
%\multicolumn{1}{c}{(16)} &
%\multicolumn{1}{c}{(17)} \\
\hline
2 &  0.01 & 2.2$\times10^6$ & -2.92 & 52\,659.745(*) & 1898.9837(*) & 1898.3517 & 4f$^3$($^4$D$^\circ$)5d $^5$F$^\circ$ & 4 & 4f$^2$($^3$H)5d($^2$D)($^4$G)6s $^5$G & 3 & 47\,404.355 & 100\,064.1 & 1.223 & 0.944 \\%& 9.2$\times10^1$ & 7.3$\times10^1$ \\
2 &  0.00 & 6.0$\times10^5$ & -3.49 & 52\,648.771(*) & 1899.3796(*) & 1898.7474 & 4f$^3$($^4$I$^\circ$)6s ($\frac{11}{2}$,$\frac{1}{2}$)$^\circ$ & 5 & 4f$^2$($^3$F)5d$^2$($^3$F) $^5$G & 5 & 32\,309.729 & 84\,958.5 & 0.875 & 1.201 \\%& 2.2$\times10^4$ & 1.3$\times10^1$ \\
2 &  0.00 & 2.9$\times10^6$ & -2.80 & 52\,645.550(*) & 1899.4958(*) & 1898.8636 & 4f$^3$($^2$H2$^\circ$)5d $^1$G$^\circ$ & 4 & 4f$^2$($^3$H)5d$^2$($^3$P) $^5$H & 5 & 36\,517.350 & 89\,162.9 & 1.067 & 1.173 \\%& 2.9$\times10^2$ & 1.4$\times10^1$ \\
2 & 0.08 & 8.0$\times10^7$ & -1.36 & 52\,643.587(*) & 1899.5666(*) & 1898.9344 & 4f$^3$($^4$F$^\circ$)5d $^3$H$^\circ$ & 4 & 4f$^3$($^2$D1$^\circ$)6p $^3$F & 3 & 32\,077.113 & 84\,720.7 & 0.948 & 1.076 \\%& 8.0$\times10^2$ & 3.7$\times10^0$ \\
2 & 0.01 & 4.4$\times10^6$ & -2.62 & 52\,636.914(*) & 1899.8074(*) & 1899.1752 & 4f$^3$($^2$H2$^\circ$)5d $^3$K$^\circ$ & 6 & 4f$^2$($^3$H)5d$^2$($^3$F) $^5$H & 7 & 29\,643.186 & 82\,280.1 & 0.990 & 1.133 \\%& 6.5$\times10^2$ & 7.4$\times10^0$ \\
\multicolumn{1}{c}{\vdots}&\multicolumn{1}{c}{\vdots}&\multicolumn{1}{c}{\vdots}&\multicolumn{1}{c}{\vdots}&\multicolumn{1}{c}{\vdots}&\multicolumn{1}{c}{\vdots}&\multicolumn{1}{c}{\vdots}&\multicolumn{1}{c}{\vdots}&\multicolumn{1}{c}{\vdots}&\multicolumn{1}{c}{\vdots}&\multicolumn{1}{c}{\vdots}&\multicolumn{1}{c}{\vdots}&\multicolumn{1}{c}{\vdots}&\multicolumn{1}{c}{\vdots}&\multicolumn{1}{c}{\vdots} \\
2 & 0.12 & 1.1$\times10^6$ & -2.44 & 21\,331.734(*) & 4687.8514(*) & 4686.5398 & 4f$^3$($^4$I$^\circ$)6d $^3$H$^\circ$ & 5 & 4f$^3$($^4$F$^\circ$)5f $^3$G & 5 & 108\,852.966 & 130\,184.7 & 1.087 & 1.165 \\%& 4.4$\times10^2$ & 1.5$\times10^3$ \\
1 & 0.00 & 2.3$\times10^3$ & -5.12 & 21\,328.712(*) & 4688.5156(*) & 4687.2038 & 4f$^3$($^2$H2$^\circ$)6s $^3$H$^\circ$ & 5 & 4f$^3$($^4$I$^\circ$)6p $^3$I & 5 & 43\,022.1 & 64\,350.812 & 1.129 & 0.849 \\%& 7.5$\times10^3$ & 1.3$\times10^0$ \\
R & 0.01 & 2.2$\times10^5$ & -3.14 & 21\,325.613(800) & 4689.1970(1800) & 4687.8850 & 4f$^3$($^4$I$^\circ$)6d $^3$I$^\circ$ & 6 & 4f$^3$($^4$F$^\circ$)5f $^5$I & 6 & 105\,866.787 & 127\,192.4 & 0.966 & 1.052 \\%& 1.2$\times10^3$ & 2.0$\times10^3$ \\
L & 0.11 & 4.5$\times10^6$ & -1.83 & 21\,320.284(6) & 4690.3690(13) & 4689.0567 & 4f$^4$ $^5$I & 7 & 4f$^3$($^4$I$^\circ$)5d $^3$H$^\circ$ & 6 & 3714.548 & 25\,034.832 & 1.177 & 1.241 \\%& 1$\times10^{27}$ & 1.4$\times10^3$ \\
2 & 0.01 & 1.1$\times10^4$ & -4.44 & 21\,319.013(*) & 4690.6487(*) & 4689.3363 & 4f$^3$($^4$I$^\circ$)6d $^5$H$^\circ$ & 4 & 4f$^2$($^1$I)5d($^2$D)($^2$H)6s $^3$H & 5 & 106265.687 & 127584.7 & 0.993 & 1.108 \\%& 4.3$\times10^2$ & 1.1$\times10^5$ \\
\multicolumn{1}{c}{\vdots}&\multicolumn{1}{c}{\vdots}&\multicolumn{1}{c}{\vdots}&\multicolumn{1}{c}{\vdots}&\multicolumn{1}{c}{\vdots}&\multicolumn{1}{c}{\vdots}&\multicolumn{1}{c}{\vdots}&\multicolumn{1}{c}{\vdots}&\multicolumn{1}{c}{\vdots}&\multicolumn{1}{c}{\vdots}&\multicolumn{1}{c}{\vdots}&\multicolumn{1}{c}{\vdots}&\multicolumn{1}{c}{\vdots}&\multicolumn{1}{c}{\vdots}&\multicolumn{1}{c}{\vdots} \\
12 & 0.10 & 8.6$\times10^3$ & -2.55 & 2126.8(*) & 47\,019.00(*) & 47\,006.18 & 4f$^3$($^2$K$^\circ$)5d $^1$H$^\circ$ & 5 & 4f$^4$ $^1$G3 & 4 & 46\,408.0 & 48\,534.8 & 1.014 & 1.005 \\%& 3.1$\times10^4$ & 3.1$\times10^4$ \\
1  & 0.03 & 1.4$\times10^3$ & -3.32 & 2089.525(*) & 47\,857.7667(*) & 47\,844.7207 & 4f$^4$ $^3$F2 & 3 & 4f$^3$($^4$I$^\circ$)5d $^3$G$^\circ$ & 4 & 24\,050.7 & 26\,140.225 & 1.113 & 1.060 \\%& 1.0$\times10^{27}$ & 1.6$\times10^4$ \\
1  & 0.05 & 1.3$\times10^3$ & -3.34 & 2059.925(*) & 48\,545.4568(*) & 48\,532.2233 & 4f$^4$ $^3$M & 8 & 4f$^3$($^4$I$^\circ$)5d $^3$I$^\circ$ & 7 & 21\,943.3 & 24\,003.225 & 0.948 & 1.171 \\%& 3.8$\times10^6$ & 3.6$\times10^4$ \\
1 & 0.07 & 1.6$\times10^3$ & -3.23 & 2007.711(*) & 49\,807.9654(*) & 49\,794.3880 & 4f$^4$ $^3$I1 & 7 & 4f$^3$($^4$I$^\circ$)5d $^3$K$^\circ$ & 8 & 25\,434.2 & 27\,441.911 & 1.137 & 1.138 \\%& 3.8$\times10^6$ & 5.6$\times10^3$ \\
12 & 0.02 & 1.8$\times10^3$ & -3.17 & 2005.9(*) & 49\,852.93(*) & 49\,839.34 & 4f$^4$ $^3$H3 & 5 & 4f$^3$($^2$G1$^\circ$)5d $^3$H$^\circ$ & 5 & 38\,143.2 & 40\,149.1 & 1.028 & 1.120 \\% & 4.7$\times10^4$ & 1.0$\times10^4$ \\
\hline
\end{tabular}
\tablefoot{The electronic version of this table is available at the CDS. In Col. 1, the label `1' indicates (experimentally) unknown lower level, `2' indicates unknown upper level, `12' indicates that both levels are unknown, `L' indicates a transition observed in the laboratory, and `R' indicates a transition with known levels but unobserved in the laboratory. The remaining columns are: (2) cancellation factor as defined in \citep{cowan1981theory}, (3)--(4) weighted TP and log of the weighted (absorption) oscillator strength, where $g_u$ and $g_l$ refer to statistical weights of the upper and lower energy levels, respectively, (5)--(7) vacuum Ritz wavenumber, vacuum Ritz wavelength, and air Ritz wavelength converted using the three term dispersion formula from \citep{peck1972dispersion}, respectively, where uncertainties are given in parentheses in units of the final decimal places for lines with types `L' and `R', otherwise, `*' indicates at least one unknown energy level for a transition, (8)--(13) level term labels, $J$ values, and energies, where the term labels of unknown levels are from the leading eigenvectors in the $LS$ coupling scheme, and (14)--(15) upper and lower level Land\'{e} $g$-factors.}
\end{sidewaystable*} %800--1200~\AA{} and $g_uA>10^7$ for the 4f$^3$5d -- 4f$^3$5f transitions with level energies below 131\,000~cm$^{-1}$ (even) and 47\,500~cm$^{-1}$ (odd),
The wavelength range (1900--50\,000~\AA{}) is selected for the applicability of the data in vacuum-UV--mid-IR spectroscopic studies. Based on line strength significance and accuracy, the range of transitions presented is as follows: 1900--50\,000~\AA{} and $g_uA>10^3$~s$^{-1}$ for transitions of the 4f$^3$(4f, 5d, 6s, and 6p) configurations with even parity level energies below 131\,000~cm$^{-1}$ and odd parity below 47\,500~cm$^{-1}$ (calculated under case 1 of Table \ref{tab:CI_space}), 2060--2950~\AA{} and $g_uA>10^3$~s$^{-1}$ for transitions with upper levels from the 4f$^3$(7s + 6d) configurations with even parity level energies below 80\,700~cm$^{-1}$ and odd parity level energies within 103\,000--112\,000~cm$^{-1}$ (calculated under case 2 of Table \ref{tab:CI_space}), and 3700--30\,000~\AA{} and $g_uA>10^4$~s$^{-1}$ for transitions with lower levels from the 4f$^3$(7s + 6d) configurations with even parity level energies below 131\,000~cm$^{-1}$ and odd parity level energies within 103\,000--112\,000~cm$^{-1}$ (calculated under case 2 of Table \ref{tab:CI_space}). 

%\begin{table}[]
%    \caption{Nd abundances ($\log\varepsilon_{\rm Nd}$) of HD~170973 and HD~144893 determined using calculated TPs of Nd~III and experimental TPs of Nd~II.}
%    \label{tab:CI_space}
%    \centering
%    \renewcommand{\arraystretch}{1.3}
%    \begin{tabular}{lcc}
%    \hline\hline
         %& \multicolumn{2}{c}{$\log\varepsilon_{\rm Nd}$}\\
 %        \multicolumn{1}{c}{TP} & HD~170973 & HD~144893 \\ \hline
 %        Nd~III (present work) & & \\
 %        Nd~III \citep{gaigalas2019extended} & & \\
 %        Nd~II \citep{den2003improved} & & \\\hline
 %   \end{tabular}
%\end{table}

The uncertainties in the calculated TPs and branching ratios are indicative from the cancellation factors \citep{cowan1981theory} in Table \ref{tab:nd_tps}. Smaller cancellation factors indicate larger uncertainties due to larger destructive interference in the summations of the dipole matrix element evaluation. For context, around 90\% of the transitions of Nd~III observed in the laboratory are calculated to have cancellation factors greater than 0.05. The calculated branching ratios of these lines were compared with their experimental relative intensities in their classifications \citep{ding2023spectrum}. 

Level lifetimes of the 4f$^3$5d configuration calculated in the present work from additionally considering 5p-excited configurations are on average 60\% larger compared to those calculated in \cite{ding2023spectrum}. This improved the mean ratio between the five experimental lifetimes measured by \cite{zhang2002measurement} and their calculated values from 1.4 to 0.9. However, the root-mean-square deviation from the five experimental values increased from 61 to 88~ns%by 50\%
. Furthermore, the changes in lifetime for levels of other configurations (not 4f$^3$5d) were less than 10\% when 5p-excited configuration interactions of the present work are included, indicated by transitions with $g_uA > 10^7$~s$^{-1}$. To reduce computational complexity, the 5p$^6$4f$^3$5f $\leftrightarrow$ 5p$^5$4f$^4$5f configuration interaction was not included due to negligible changes observed for the 4f$^3$5f lifetimes when included. We expect more lifetime measurements for Nd~III levels to be required to conclude whether the present lifetime calculations provide improvements compared to those made in \cite{ding2023spectrum}. 

We would also like to note that the lifetimes calculated in the present work in Table \ref{tab:nd_levels} are only from electric dipole (E1) transitions, so lifetimes are not listed for levels with no lower level of opposite parity or $J$ value within $\pm1$, e.g., the lowest-lying 4f$^4$ levels labelled with $^5$I, $^5$F, and $^3$K2 terms.

\subsection{Implications for the Nd III energy level analysis}
The implications of the new calculations including 5p-excited configuration interactions on all experimental identification of Nd~III energy levels were investigated. The lowest levels of the 5p-excited configurations were estimated to lie several 10\,000~cm$^{-1}$ above the highest-lying known level. Therefore, changes to eigenvector compositions and branching fractions of known levels were very small. None of the five leading eigenvector components of the levels of Table~\ref{tab:nd_levels} belong to the 5p-excited configurations. Experimental energy level identifications using calculations without 5p-excited configuration interactions \citep[Table~\ref{tab:nd_levels} and ][]{ding2023spectrum} agree with calculations from the present work including 5p-excited configuration interactions. In a few cases, branching fractions obtained in the new calculations with 5p-excited configuration interactions agreed better with experimental relative intensities. Only two of the highest-lying 4f$^3$6p levels reported in \cite{ding2023spectrum}, 77\,231.115~cm$^{-1}$ $J=5$ and 80\,593.604~cm$^{-1}$ $J=6$, are now considered uncertain and tentative due to results of the new calculations. This does not affect the optimised energies of other levels as the energies of these two levels are defined by one line each. %, their two lines in Table 6 and seven lines in Table 7 of \cite{ding2023spectrum} should also now be considered as unclassified. We note that since the two 4f$^3$6p levels were optimised using one FT line each, their false classifications do not affect the optimised level energies of other levels.

\subsection{Evaluation using Nd-rich stellar spectra}
Nd~III TPs calculated in the present work offer improvements to those from \cite{gaigalas2019extended} and \cite{ding2023spectrum}. We deduced this by comparing Nd abundance determined using 4f$^4$ -- 4f$^3$5d transitions in the Nd-rich star HD~170973 with values obtained using experimental Nd~II TPs \citep{den2003improved}. The comparison is shown in Fig.~\ref{fig3}. The analysed Nd~II and Nd~III spectral lines are listed in Table~\ref{tab:nd2_abun} and Table~\ref{tab:nd3_abun}, respectively, where $\log\varepsilon_{\rm Nd}=\log(N_{\rm Nd}/N_{\rm tot})$ + 12.04 for number densities of neodymium $N_{\rm Nd}$ and of the total number of atoms $N_{\rm tot}$. Abundances were derived by fitting synthetic Nd II-III line profiles to observed features using a model atmosphere with effective temperature $T_{\rm eff}=11200$~K and surface gravity $\log g$=3.8 \citep{2011mast.conf...69R}. According to the analysis of rare-earth abundances in Ap star atmospheres as a function of effective temperature \citep[][see Fig. 2 of their paper]{2017AstL...43..252R}, Nd ionisation equilibrium is observed in the temperature range 10\,000 -- 12\,000~K. Therefore we may expect Nd abundance from Nd II and Nd III lines to agree for HD~170973, which we have observed with the present TP calculations of this work. 
%It means that Cownan's calculations for Nd III provide TP scale consistent with %the experimental set of Nd II TP's.    
We note that $\log\varepsilon_{\rm Nd}$ obtained using TPs of \cite{ding2023spectrum}, calculated without considering 5p-excitation (not shown), are about 0.2~dex lower, indicating the improvements of the present calculations. 
\begin{figure}
    \centering
    \includegraphics[width=\linewidth]{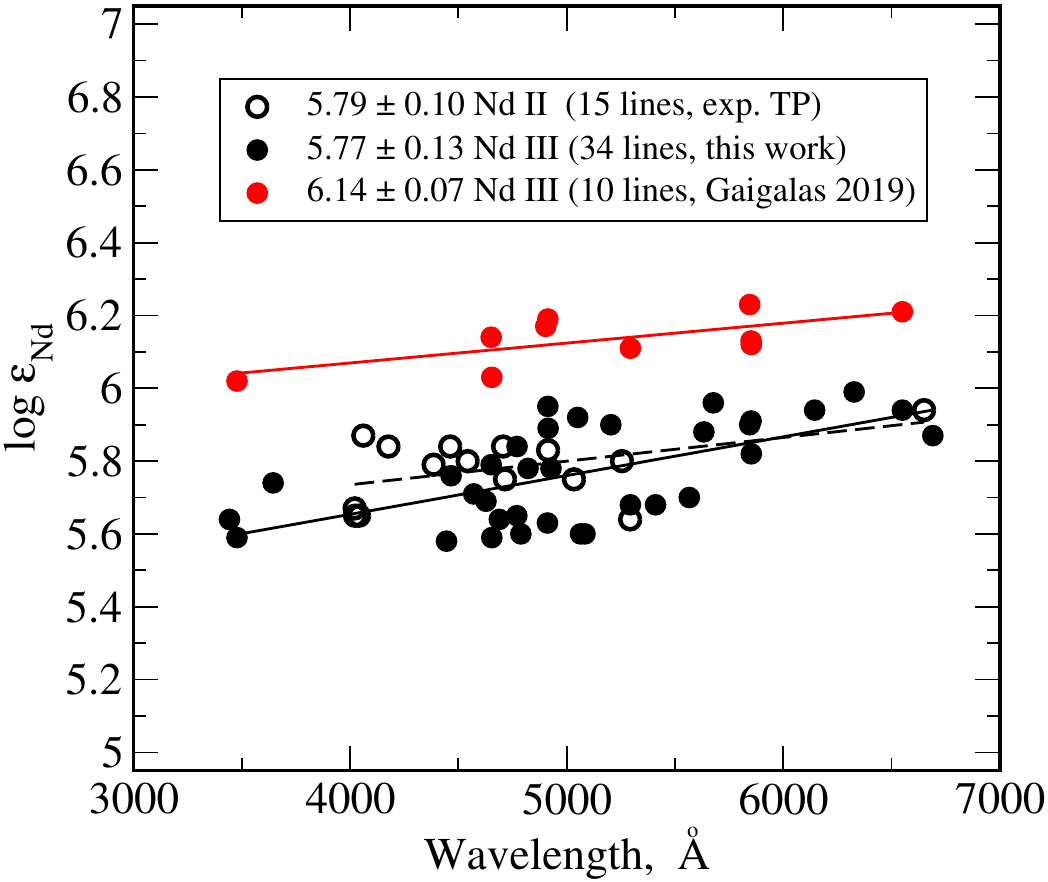}
    \caption{Nd abundance $\log\varepsilon_{\rm Nd}$ in the star HD~170973 against wavelength determined using experimental Nd~II TPs \citep[open black,][]{den2003improved} and Nd~III TPs of this work (filled black circles) and of \cite{gaigalas2019extended} (filled red circles). Mean values and their uncertainties are in the legend. Linear regressions are shown by solid and dashed lines.}
    \label{fig3}
\end{figure}
\begin{table}[]
    \caption{Nd II spectral lines used for Nd abundance determination of HD~170973.}
    \label{tab:nd2_abun}
    \centering
    \renewcommand{\arraystretch}{1.3}
    \begin{tabular}{cccccc}
    \hline\hline
    $\lambda^{\text{air}}$ & $E_l$ & $\log(g_lf)$ & $\log\varepsilon_{\rm Nd}$ \\
    (\AA{}) & (eV) & &\\
    \hline
  4021.327 & 0.321 & -0.10 & 5.67 \\
  4023.000 & 0.559 & 0.04 & 5.65 \\
  4041.056 & 0.471 & -0.53 & 5.65 \\
  4061.080 & 0.471 & 0.55 & 5.87 \\
  4177.320 & 0.064 & -0.10 & 5.84 \\
  4385.661 & 0.205 & -0.30 & 5.79 \\
  4462.409 & 0.205 & -0.91 & 5.84 \\
  4542.600 & 0.742 & -0.28 & 5.80 \\
  4706.543 & 0.000 & -0.71 & 5.84 \\
  4715.586 & 0.205 & -0.90 & 5.75 \\
  4914.382 & 0.380 & -0.70 & 5.83 \\
  5033.507 & 1.136 & -0.47 & 5.75 \\
  5255.506 & 0.205 & -0.67 & 5.84 \\
  5293.163 & 0.823 & 0.10 & 5.64 \\
  6650.517 & 1.953 & -0.11 & 5.94 \\
    \hline    \hline
    \end{tabular}
\tablefoot{Using air wavelength $\lambda^{\text{air}}$, lower level energy $E_l$, and $\log(g_lf)$ from \cite{den2003improved}.}
\end{table}
\begin{table}[]
    \caption{Nd III spectral lines used for Nd abundance determination of HD~170973.}
    \label{tab:nd3_abun}
    \centering
    \renewcommand{\arraystretch}{1.3}
    \begin{tabular}{cccccc}
    \hline\hline
    $\lambda^{\text{air}}$ & $E_l$ & $\log(g_lf)$ & $\log\varepsilon_{\rm Nd}$ &  $\log(g_lf)$ & $\log\varepsilon_{\rm Nd}$ \\
    (\AA{}) & (eV) & \multicolumn{2}{c}{(This work)}& \multicolumn{2}{c}{\citep{gaigalas2019extended}}\\
    \hline
      3442.7834 &  0.141 & -1.73 &  5.64   &&                  \\
      3477.8330 &  0.000 & -1.85 &  5.59   &-2.31 & 6.02     \\
      3644.3510 &  0.461 & -0.78 &  5.74   &&                  \\
      4445.0012 &  0.296 & -2.33 &  5.58   &&                  \\
      4466.3512 &  0.461 & -2.54 &  5.76   &&                  \\
      4570.6374 &  0.141 & -2.14 &  5.71   &&                  \\
      4627.2556 &  0.461 & -2.38 &  5.69   &&                  \\
      4651.6172 &  0.000 & -2.07 &  5.79   &-2.39 & 6.11     \\
      4654.3195 &  0.000 & -1.94 &  5.59   &-2.45 & 6.03     \\
      4689.0565 &  0.461 & -1.83 &  5.64   &&                  \\
      4769.6184 &  0.296 & -1.86 &  5.65   &&                  \\
      4770.8900 &  1.510 & -1.72 &  5.84   &&                  \\
      4788.4584 &  0.461 & -1.92 &  5.60   &&                  \\
      4821.9906 &  0.296 & -2.78 &  5.78   &&                  \\
      4903.2380 &  0.000 & -2.53 &  5.89   &-3.36 & 6.17     \\
      4911.6527 &  0.141 & -1.74 &  5.63   &&                  \\
      4912.9436 &  0.000 & -1.94 &  5.95   &-2.32 & 6.19     \\
      4914.0941 &  0.461 & -1.27 &  5.89   &&                  \\
      4927.4877 &  0.461 & -1.00 &  5.78   &&                  \\
      5050.6952 &  0.296 & -1.24 &  5.92   &&                  \\
      5084.6597 &  0.141 & -2.65 &  5.60   &&                  \\
      5203.9236 &  0.141 & -0.83 &  5.90   &&                  \\
      5294.1133 &  0.000 & -0.85 &  5.68   &-1.19 & 6.11     \\
      5410.0994 &  0.141 & -1.62 &  5.68   &&                  \\
      5566.0154 &  0.296 & -2.46 &  5.70   &&                  \\
      5633.5540 &  0.141 & -2.29 &  5.88   &&                  \\
      5677.1788 &  0.631 & -1.62 &  5.96   &&                  \\
      5845.0201 &  0.631 & -1.32 &  5.90   &-1.61 &  6.23     \\
      5851.5419 &  0.461 & -1.69 &  5.91   &-1.91 &  6.13     \\
      5852.4312 &  1.744 & -1.60 &  5.82   &-1.90 &  6.12     \\
      6145.0677 &  0.296 & -1.48 &  5.94   &&                  \\
      6327.2649 &  0.141 & -1.55 &  5.99   &&                  \\
      6550.2242 &  0.000 & -1.64 &  5.94   &-1.97 & 6.21     \\
      6690.8302 &  0.461 & -2.57 &  5.87   &&                  \\
    \hline
    \end{tabular}
\end{table}

%Only Nd~III transitions to the ground level from \cite{gaigalas2019extended} were used in the abundance determinations of HD~170973. These transitions are expected to be most accurately predicted and are non-ambiguous. For example, we encountered two levels predicted in \cite{gaigalas2019extended} labelled as 4f$^3$5d($^4$F$^\circ$)5d $^5$H$^\circ$$_6$, giving two $\log(gf)$s for the experimentally known transition 4f$^4$ $^3$K2$_6$ -- 4f$^3$5d($^4$F$^\circ$)5d $^5$H$^\circ$$_6$ at 5852.43~\AA{}. 
The difficulty in using the Nd~III TP data calculated by \cite{gaigalas2019extended} for the interpretation of the stellar spectra should be mentioned. Many calculated levels in the \cite{gaigalas2019extended} dataset have been labelled with the same names. For example, in the $J=6$ matrix, the 20777.62 and 21844.68~cm$^{-1}$ levels have (in their notations) the 4f(3)4I1.5d\_5I label, the 25161.85 and 26820.84~cm$^{-1}$ levels have the 4f(3)4I1.5d\_5G label, the 32640.91 and 32961.13~cm$^{-1}$ levels are attributed to the 4f(3)4F1.5d\_5H label, and so on. We can illustrate impacts of such ambiguity in the interpretation of the HD~170973 stellar spectrum using an example of the experimentally known transition 4f$^4$ $^3$K2$_6$ -- 4f$^3$5d($^4$F$^\circ$)5d $^5$H$^\circ$$_6$ at 5852.43~\AA{}. As is expected, the calculated $\log(gf)$ values are different for the transitions from the two levels of the same label as given in \cite{gaigalas2019extended}. The differences in spectral syntheses between these two $\log(gf)$s are shown by blue and dashed blue lines in Fig.~\ref{fig4}. The $\log(gf)$ used for the blue line appears to be the correct value, agreeing with the average abundance deviation. However, the $\log(gf)$ used for the dashed blue line belongs to the predicted level that better resembled the known 4f$^3$5d($^4$F$^\circ$)5d $^5$H$^\circ$$_6$ level at 31146.457~cm$^{-1}$, based on relative intensity analysis of the stronger observed and predicted lines to the ground term levels and from 4f$^3$($^4$F$^\circ$)6p $^5$G$_5$. All other Nd~III transitions from \cite{gaigalas2019extended} used in the abundance determinations of HD~170973 shown in Fig. \ref{fig3} were to the ground term levels. These transitions are expected to not have ambiguous labels and hence comparable with our results. Therefore, we highlight the advantages of the present work in individual line position and stellar spectra analyses. Although in the context of kilonova opacity calculations, the difference between TPs of the present work and those from \cite{gaigalas2019extended} may not be significant, we do however expect the newly established Ritz wavelengths to allow improvements in any kilonovae spectral feature determinations.
\begin{figure*}
    \centering
    \includegraphics[width=\linewidth]{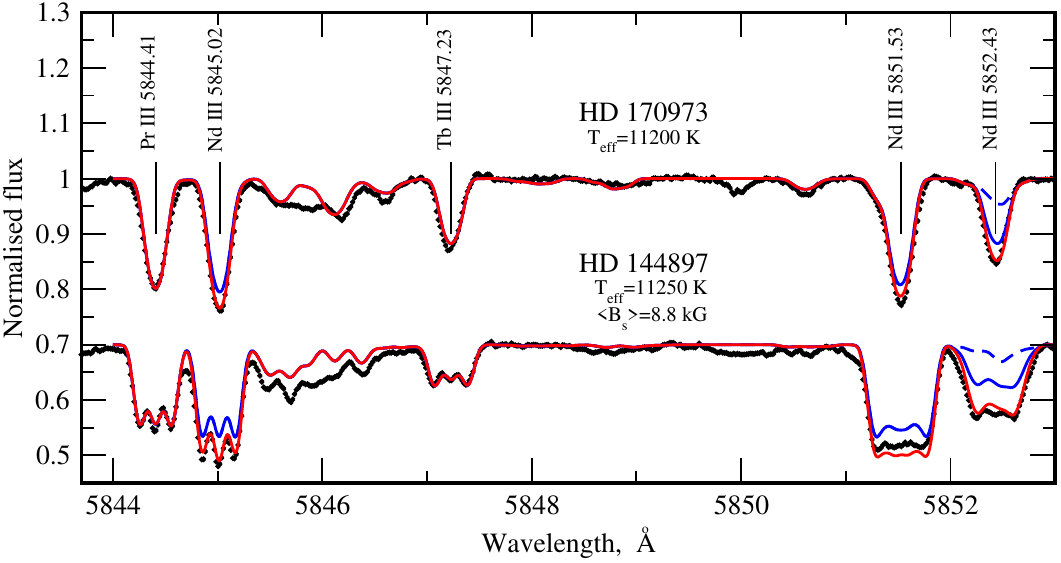}
    \caption{Observed (black filled circles) spectra in the interval 5844--5853~\AA{} of stars HD~170973 and HD~144893 with surface magnetic fields \bs$<$1~kG and \bs=8.8~kG, respectively, the coloured lines show fits to the observed spectra using Nd abundances $\log\varepsilon_{\rm Nd}=$ 5.84 and 5.77, respectively. Fits using TPs calculated in the present work are in red and fits using TPs from \cite{gaigalas2019extended} are in blue and dashed blue. All Land\'{e} $g$-factors used are calculated from the present work.}
    \label{fig4}
\end{figure*}

\section{Revision of Nd III ionisation energy}
The ionisation energy of Nd~III is the energy separation between the 4f$^4$~$^5$I$_4$ ground energy level of Nd~III and the 4f$^3$~$^4$I$^\circ$$_{9/2}$ ground energy level of Nd~IV, which can be estimated when the four levels of the 4f$^3$($^4$I$^\circ$$_{9/2}$)6s and 4f$^3$($^4$I$^\circ$$_{9/2}$)7s sub-configurations of Nd~III are known \citep{sugar1973ionization}. The method is based on solving the equation of the difference between the center-of-gravity (COG) binding energies of the 4f$^3$6s and 4f$^3$7s configurations using the Rydberg-Ritz formula,
\begin{equation}
    \Delta T=\frac{R_{\text{Nd}}Z_{\text{C}}}{[n^*(\text{6s})] ^ 2} - \frac{R_{\text{Nd}}Z_{\text{C}}}{[n^*(\text{6s}) + \Delta n^*(\text{7s}-\text{6s})] ^ 2},
\end{equation} 
where $R_{\text{Nd}}=109\,736.9$~cm$^{-1}$ is the Nd Rydberg constant, $Z_{\text{C}}=3$ is the Nd~III core charge, $n^*(n\text{s})$ is the effective quantum number of the configuration $n$s, and $\Delta n^*(\text{7s}-\text{6s})=n^*(\text{7s})-n^*(\text{6s})$. The $\Delta T$ of Nd~III, defined for the full 6s and 7s configurations, is well represented by the difference between COG energies the lowest-lying terms of the 4f$^3$($^4$I$^\circ$$_{9/2}$)6s and 4f$^3$($^4$I$^\circ$$_{9/2}$)7s sub-configurations \citep{sugar1965ionization, reader1966ionization, sugar1973ionization}. 
%\textbf{i.e., perturbations to the residual electrostatic and spin-orbit interactions between the 4f$^3$ electrons from the s electrons are negligible, which is not entirely true and evident from the small amounts of mixing in the predicted eigenvector compositions of 4f$^3$($^4$I$^\circ$$_{9/2}$)6s/7s. However, this neglected perturbation is part of $\Delta n^*(\text{7s}-\text{6s})$ and the uncertainty of $\Delta n^*(\text{7s}-\text{6s})$ therefore limits the accuracy of the method.} 
Thus, with knowledge of the quantities $\Delta T$ and $\Delta n^*(\text{7s}-\text{6s})$, $n^*(\text{6s})$ and hence the COG binding energy of 4f$^3$($^4$I$^\circ$$_{9/2}$)6s is solved, which is added to the COG energy of 4f$^3$($^4$I$^\circ$$_{9/2}$)6s relative to the ground state energy to yield the ionisation energy.

In 1973, the COG energies of the 4f$^3$($^4$I$^\circ$$_{9/2}$)6s and 4f$^3$($^4$I$^\circ$$_{9/2}$)7s sub-configurations were unknown. These quantities, as well as $\Delta n^*(\text{7s}-\text{6s})$, were interpolated by \cite{sugar1973ionization} from studies of some second and third spectra of the lanthanide atoms. With the accepted quantity $\Delta n^*(\text{7s}-\text{6s})=1.048\pm0.002$, the ionisation energy was estimated in 1973 as $178\,600\pm2400$~cm$^{-1}$.

Recently, \cite{johnson2017lanthanideI3} updated the Nd~III ionisation energy to be $178\,140\pm600$~cm$^{-1}$ using a framework of the method by \cite{sugar1973ionization}. The levels of the 4f$^3$($^4$I$^\circ$$_{9/2}$)6s and 4f$^3$($^4$I$^\circ$$_{9/2}$)7s sub-configurations were still unknown. However, improved interpolated values of $\Delta T$ and $\Delta n^*(\text{7s}-\text{6s})$ were obtained from the availability of new data since 1973 for the other neighbouring ions.

In \cite{ding2023spectrum} and the present work, the COG energies of the 4f$^3$($^4$I$^\circ$$_{9/2}$)6s and 4f$^3$($^4$I$^\circ$$_{9/2}$)7s sub-configurations were determined at 30\,269~cm$^{-1}$ and 103\,223~cm$^{-1}$ respectively, this corresponds to $\Delta T = 72\,954$~cm$^{-1}$ and has now permitted a further update of the Nd~III ionisation energy. Together with the estimated value of $\Delta n^*(\text{7s}-\text{6s})=1.0476\pm0.0052$ by \cite{johnson2017lanthanideI3}, 2.5850 is now obtained for $n^*({\text{6s}})$. As a result, the ionisation energy of Nd~III is estimated to be $178\,070\pm330$~cm$^{-1}$, in close agreement with \cite{johnson2017lanthanideI3} but with about two times higher accuracy. The uncertainty $\pm330$~cm$^{-1}$ is at two standard deviations and is fully determined by the $\pm0.0052$ uncertainty of the $\Delta n^*(\text{7s}-\text{6s})$ value estimated by \cite{johnson2017lanthanideI2}.
\section{Outlook}
This work has made significant progress with new levels of the singly excited configurations of Nd~III. Energy levels of the doubly-excited configurations of Nd~III (e.g. 4f$^2$5d$^2$, 4f$^2$5d6s, and 4f$^2$5d6p), however, remain experimentally unknown. Experimentally determining their level energies should further improve atomic structure and TP calculations of Nd~III as these overlap greatly with the configurations of known levels. More than half of the lines measured (observed) in the FT and grating spectra remain unclassified, this number is of order $10^4$. Therefore, the accurate experimental knowledge of the spectra of Nd in its first few ionisation stages is still appreciably incomplete. Further analyses of these spectra are expected to be more challenging compared to the presented work and the work of \cite{ding2023spectrum}, and more extensive experimental measurements are also likely required.

\section{Summary}
Fourier transform spectroscopy of Nd-Ar Penning and hollow cathode discharge lamps, grating spectroscopy of Nd vacuum sliding sparks, and the parameterised Cowan code atomic structure and TP calculations have enabled the first experimental classification of 355 transitions and establishment of 116 new energy levels of the 4f$^3$7s, 4f$^3$6d, and 4f$^3$5f configurations of Nd~III. One previously unknown level of the 4f$^3$5d configuration is also now identified. Together with results from \cite{ding2023spectrum}, 261 energy levels and 972 transitions of Nd III have been observed, classified, and measured at accuracies up to a few parts in 10$^8$. The experimentally determined energies of the four levels of the 4f$^3$($^4$I$^\circ$$_{9/2}$)6s and 4f$^3$($^4$I$^\circ$$_{9/2}$)7s sub-configurations enabled a revision of the Nd~III ionisation energy at $178\,070\pm330$~cm$^{-1}$. Calculated level energies, eigenvector compositions, and lifetimes are presented for energies within 15\,158--47\,500~cm$^{-1}$ and 103\,085--111\,165~cm$^{-1}$ for the odd parity system and within 0--130\,936~cm$^{-1}$ for the even parity system. Transition wavelengths with significant calculated TPs within 1900--50\,000~\AA{} are also presented and evaluated against stellar spectra. These new data will aid future experimental and theoretical investigations of the Nd~III spectrum and allow more accurate and reliable modeling of astrophysical plasmas containing Nd~III, such as in stellar atmospheres and kilonovae.

%% project  For this sample we use BibTeX plus aasjournals.bst to generate the
%% the bibliography. The sample631.bib file was populated from ADS. To
%% get the citations to show in the compiled file do the following:
%%
%% pdflatex sample631.tex
%% bibtext sample631
%% pdflatex sample631.tex
%% pdflatex sample631.tex
\begin{acknowledgements}
    This work was supported at Imperial College by the STFC of the UK, grant numbers ST/S000372/1, ST/N000838/1, and ST/W000989/1, the Bequest of Prof. Edward Steers, and at the Institute of Spectroscopy of the Russian Academy of Sciences the research project FFUU-2022-0005. We are grateful to Prof. C. R. Cowley for sharing the unpublished Nd~III line lists of Dr H. M. Crosswhite. We also thank Dr J.-F. Wyart and Prof. W.-\"{U}. Tchang-Brillet for providing the Nd vacuum sliding spark grating plates recorded at NIST, and we acknowledge Dr N. Spector and Dr J. Sugar for the recording of these plates.
\end{acknowledgements}

%------------------------------------------------

\bibliographystyle{aa}
\bibliography{bibliography}

\end{document}